\documentclass[aps,prl,english,amsmath,amsfonts,amssymb,superscriptaddress,twocolumn,showpacs,citeautoscript,reprint,longbibliography]{revtex4-1}
\usepackage[T1]{fontenc}
\usepackage[latin9]{inputenc}
\usepackage{amsmath}
\usepackage{amssymb}
\usepackage{wasysym}
\usepackage{esint}
\usepackage{graphicx}
\usepackage{times}
\usepackage{nicefrac}
\usepackage{appendix}
\usepackage{textcase}
\usepackage{subfigure}
\usepackage{empheq}
\usepackage{color}
\usepackage{leftidx}
\usepackage{makecell}
\usepackage{stackrel}
\usepackage{lineno}
\makeatletter
\@ifundefined{textcolor}{}
{%
 \definecolor{BLACK}{gray}{0}
 \definecolor{WHITE}{gray}{1}
 \definecolor{RED}{rgb}{1,0,0}
 \definecolor{GREEN}{rgb}{0,1,0}
 \definecolor{BLUE}{rgb}{0,0,1}
 \definecolor{CYAN}{cmyk}{1,0,0,0}
 \definecolor{MAGENTA}{cmyk}{0,1,0,0}
 \definecolor{YELLOW}{cmyk}{0,0,1,0}
}

\renewcommand{\vec}[1]{\mathbf{#1}}

\renewcommand{\b}{\beta}

\newcommand{\add}[1]{\if\a\b{{\color{red} #1}}\else{#1}\fi}

\newcommand{\bracket}[1]{\langle #1 \rangle}
\newcommand{\ket}[1]{| #1 \rangle}
\newcommand{\bra}[1]{\langle #1 |}
\newcommand{\im}{\mathrm{i}}

\renewcommand{\eqref}[1]{(\ref{eq:#1})}

\newcommand{\figref}[1]{Fig.~\ref{fig:#1}}
\newcommand{\Figref}[1]{Figure~\ref{fig:#1}}

\newcommand{\appref}[1]{Appendix~\ref{sec:#1}}

\newcommand{\trace}[1]{{\rm Tr} \left[ #1 \right]}

\makeatother
\usepackage{babel}

\begin{document}
\title{Impact of nuclear vibrations on van der Waals and Casimir
  interactions \\ at zero and finite temperature}

\author{Prashanth S. Venkataram}
\affiliation{Department of Electrical Engineering, Princeton
  University, Princeton, New Jersey 08544, USA}
\author{Jan Hermann}
\affiliation{Physics and Materials Science Research Unit, University
  of Luxembourg, L-1511 Luxembourg}
\author{Teerit J. Vongkovit }
\affiliation{Department of Physics, Princeton
  University, Princeton, New Jersey 08544, USA}
\author{Alexandre Tkatchenko}
\affiliation{Physics and Materials Science Research Unit, University
  of Luxembourg, L-1511 Luxembourg}
\author{Alejandro W. Rodriguez}
\affiliation{Department of Electrical Engineering, Princeton
  University, Princeton, New Jersey 08544, USA}

\date{\today}

\begin{abstract}
  Van der Waals (vdW) and Casimir interactions depend crucially on
  material properties and geometry, especially at molecular scales,
  and temperature can produce noticeable relative shifts in
  interaction characteristics. Despite this, common treatments of
  these interactions ignore electromagnetic retardation, atomism, or
  contributions of collective mechanical vibrations (phonons) to the
  infrared response, which can interplay with temperature in
  nontrivial ways. We present a theoretical framework for computing
  electromagnetic interactions among molecular structures, accounting
  for their geometry, electronic delocalization, short-range
  interatomic correlations, dissipation, and phonons at atomic scales,
  along with long-range electromagnetic interactions among themselves
  or in the vicinity of continuous macroscopic bodies. We find that in
  carbon allotropes, particularly fullerenes, carbyne wires, and
  graphene sheets, phonons can couple strongly with long-range
  electromagnetic fields, especially at mesoscopic scales
  (nanometers), to create delocalized phonon polaritons that
  significantly modify the infrared molecular response. These
  polaritons especially depend on the molecular dimensionality and
  dissipation, and in turn affect the vdW interaction free energies of
  these bodies above a macroscopic gold surface, producing
  nonmonotonic power laws and nontrivial temperature variations at
  nanometer separations that are within the reach of current Casimir
  force experiments.
\end{abstract}

\maketitle 

Van der Waals (vdW) interactions play an integral role in binding and
interaction energies of molecules in condensed
phases~\cite{WoodsRMP2015, Langbein1974,TkatchenkoADFM2014}, and their
long-range many-body nature~\cite{TkatchenkoJCP2013, GobreNCOMMS2013,
  DiStasioJPCM2014, TkatchenkoADFM2014, AmbrosettiSCIENCE2016} can
modify phonons in molecular crystals to the extent of producing
qualitatively different predictions of thermodynamic stability at
finite temperature compared to common pairwise
approximations~\cite{ReillyPRL2014, ReillyCS2015,
  HojaCMS2017}. However, while these treatments of many-body vdW
interactions account for multiple scattering and electromagnetic (EM)
screening to all orders and derive material properties from ab-initio
calculations, they only account for valence electrons and not phonons
in the molecular response, despite the large role of the latter in
interactions at finite temperature; moreover, these treatments neglect
electromagnetic retardation, which becomes important at length scales
where phononic contributions to molecular response become important
too. Such accounts of phonons have more typically arisen in continuum
treatments of Casimir interactions, where polarizable objects are
modeled as dipoles or continuum local dielectrics with infrared
resonances determined empirically~\cite{BuhmannPRA2012,
  RodriguezNATURE2011, ZouNATURE2013}. In the related domain of
thermal radiation, vibrational resonances have been treated
atomistically via mechanical Green's function~\cite{DharJSP2006,
  TianPRB2014, TianPRB2012, MingoPRB2003, ChiloyanNATURE2015,
  PendryPRB2016} as well as molecular dynamics~\cite{HenryPRL2008,
  EsfarjaniPRB2011, NoyaPRB2004, CuiJPCA2015} methods, but these have
the respective pitfalls of being limited to bulk materials or using
heuristic pairwise approximations to noncovalent interactions.

In this paper, we develop and apply a framework for computing
retarded, many-body (RMB) vdW interactions in mesoscopic systems,
where molecules can be treated in an ab-initio atomistic way and
larger bodies can be treated via continuum electrodynamics, to include
the impact of phonons and dissipation in molecular response as well as
finite temperature, based on related recent
developments~\cite{VenkataramPRL2017}. In particular, we focus on the
interactions of fullerenes, carbyne wires, and graphene sheets with a
gold surface, which we approximate as a perfect electrically
conducting plane for computational simplicity, and compare interaction
energies with and without phonon contributions at $T = 0$ and
$T = 300~\mathrm{K}$ to each other, as well as to predictions from
dipolar and continuum treatments where appropriate. We find that
phonons can significantly delocalize the molecular response, which is
especially relevant when the molecule is close to the surface, in a
manner strongly dependent on the molecular dimensionality, size, and
dissipation properties, due to the shape-dependent coupling of those
phonons with EM fields to form phonon polaritons. Moreover, in
contrast to common macroscopic treatments of Casimir forces in bulk
media, which find nontrivial temperature effects only at large
separations of at least 1 micron~\cite{SushkovNATURE2011}, we find
that phonon-induced nonlocality can lead to temperature-sensitive vdW
interactions even at small separations. In particular, we show that
``0-dimensional'' fullerenes retain a relatively localized response
and consequently lesser deviations of finite from zero-temperature
free energies and power laws at nanometer separations, while
``1-dimensional'' carbyne wires exhibit much larger quantitative but
also qualitative deviations, including nonmonotonic power laws. By
contrast, ``2-dimensional'' graphene sheets have many more avenues for
dissipation and stronger bonds than isolated compact molecules,
leading to damping of the nonlocal response, in which case finite
temperature effects, while larger than previously predicted, only
become evident at larger (tens of nanometers) separations. We expect
our predictions to be relevant to ongoing experiments on vdW
interactions among molecules and metallic objects at nanometric
scales~\cite{WagnerNATURE2014}.

\begin{figure}[t!]
\centering
\includegraphics[width=0.9\columnwidth]{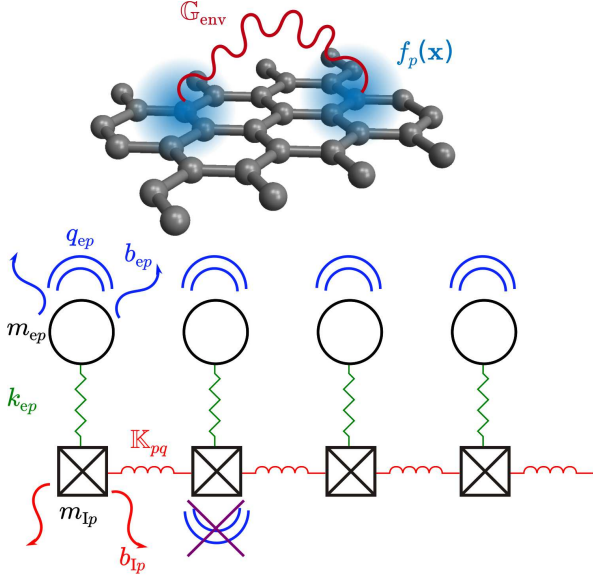}
\caption{\textbf{RMB model of molecular response.} A collection of
  atoms with electronic polarization response modeled as Gaussian
  basis functions $f_{p} (\vec{x})$ interact via long-range
  electromagnetic fields $\mathbb{G}_{\mathrm{env}}$.  The individual
  electronic response of each atom arises from the coupling of valence
  electronic and phononic excitations via short-range interactions,
  represented schematically: for every atom $p$, a nuclear oscillator
  of mass $m_{\mathrm{I}p}$ with dissipation $b_{\mathrm{I}p}$ is
  connected to nuclear oscillators of other atoms $q$ via anisotropic
  spring constants $\mathbb{K}_{pq}$, and to an electronic oscillator
  of mass $m_{\mathrm{e}p}$ with dissipation $b_{\mathrm{e}p}$ and
  isotropic spring constant $k_{\mathrm{e}p}$; only the electrons
  couple directly to long-range EM fields with effective charge
  $q_{\mathrm{e}p}$.}
\label{fig:schematic}
\end{figure}

\textbf{Method}.---At temperature $T$, the free energy of interaction
among a collection of $N_{\mathrm{mol}}$ disjoint molecules labeled
$l$, with electric susceptibilities $\mathbb{V}_{l}$, and macroscopic
bodies with a composite susceptibility $\mathbb{V}_{\mathrm{env}}$ is
given by~\cite{RahiPRD2009, ReidPRA2013, VenkataramPRL2017},
\begin{equation} \label{eq:vdWenergy} \mathcal{F} = k_{\mathrm{B}}
  T\sum_{n = 0}^{\infty} {}' \underbrace{\ln(\det(\mathbb{T}_{\infty}
  \mathbb{T}^{-1}))}_{\Phi(\im\xi_{n})}.
\end{equation}
whose integrand $\Phi(\im\xi)$ depends on the inverse of the
scattering ``T'' operator
$\mathbb{T}^{-1} = \sum_{l} \mathbb{V}_{l}^{-1} -
\mathbb{G}_{\mathrm{env}}$, which encodes EM scattering to all orders
among the molecules mediated by the continuum bodies, and
$\mathbb{T}_{\infty}^{-1} = \prod_{l} \mathbb{T}_{l\infty}^{-1}$,
which encodes the scattering properties of each molecule isolated in
vacuum,
$\mathbb{T}_{l\infty}^{-1} = \mathbb{V}_{l}^{-1} - \mathbb{G}_{0}$;
$\mathbb{G}_{\mathrm{env}}$ denotes the imaginary-frequency Green's
function of the macroscopic bodies, which solves the macroscopic
Maxwell's equations (setting $\epsilon_{0} = 1$):
\begin{equation}
  \left[\nabla \times \nabla \times + \frac{\xi^2}{c^2}(\mathbb{I} + \mathbb{V}_{\mathrm{env}})\right] \mathbb{G}_{\mathrm{env}} = \mathbb{I}.
\end{equation}
Note that all of these quantities depend on frequency, but this is
notationally suppressed for brevity. The integrand $\Phi$ is evaluated
at imaginary Matsubara frequencies
$\xi_{n} = \frac{2\pi k_{\mathrm{B}} Tn}{\hbar}$, with the prime on
the summation denoting the $n = 0$ term contributing a half-weight
relative to the other terms and the $T \to 0$ limit reducing to an
integral over all $\xi$.

While the vacuum EM Green's function
$\tensor{G}_{0} (\im\xi, \vec{x}, \vec{x}') = (\mathbb{I} -
\frac{c^{2}}{\xi^{2}} \nabla \otimes \nabla) \frac{e^{-\xi |\vec{x} -
    \vec{x}'|/c}}{4\pi |\vec{x} - \vec{x}'|}$ is known analytically,
$\mathbb{G}_{\mathrm{env}}$ and $\mathbb{V}_{l}$ typically must be
constructed numerically. The former can be done using one of many
available classical EM techniques~\cite{Johnson2011, WoodsRMP2015,
  RodriguezADP2015, RodriguezNATURE2011}. The latter in principle
requires descriptions accounting for the quantum delocalization and
transitions of electrons, and while the susceptibilities
$\mathbb{V}_{l}$ are basis-independent quantities, computational
treatment of electromagnetic interactions between arbitrarily
delocalized electrons becomes challenging. However, recent work in the
context of vdW interactions~\cite{TkatchenkoJCP2013, DiStasioJPCM2014,
  TkatchenkoADFM2014, AmbrosettiJCP2014} has demonstrated accurate
results by expressing the molecular susceptibilities,
\begin{equation}
  \mathbb{V}_{l} = -\frac{\xi^2}{c^2}\sum_{p,i,q,j} \alpha_{pi,qj} \ket{\vec{f}_{pi}}
  \bra{\vec{f}_{qj}},
\end{equation}
in terms of localized basis functions $\ket{\vec{f}_{pi}}$
representing the EM response of valence electrons via dipolar ground
state wavefunctions of effective polarizabilities $\alpha$. We briefly
describe the construction of the molecular response functions
$\mathbb{V}_{l}$ and their underlying assumptions, shown schematically
in~\figref{schematic}, and leave more detailed descriptions
to~\appref{RMBresponse}. For the insulating and weakly metallic
molecular systems we consider, the ground state electron density is
relatively localized around each atom, so once it is found via DFT, it
can be partitioned into atomic fragments that incorporate short-range
quantum exchange, correlation, hybridization, and electrostatic
effects. These fragments are then mapped onto a set of harmonic
oscillators for each atom $p$ in each molecule, namely a single
effective electronic oscillator of charge $q_{\mathrm{e}p}$, mass
$m_{\mathrm{e}p}$, spring constant $k_{\mathrm{e}p}$ connecting only
to the nucleus, and damping coefficient $b_{\mathrm{e}p}$,
representing the valence electron, and a single nuclear oscillator of
mass $m_{\mathrm{I}p}$, anisotropic spring constants $\mathbb{K}_{pq}$
connecting to other nuclei, and damping coefficient $b_{\mathrm{I}p}$,
representing the nucleus screened by inner electrons which give rise
to phonons.

Within the RMB framework, only the \emph{valence} electronic
oscillators are assumed to couple to long-range EM fields via their
charges $q_{\mathrm{e}p}$, thereby giving rise to the molecular
response in the first place, but the features of the response are
strongly influenced by the coupling of the electronic oscillators to
their corresponding mobile nuclear oscillators via $k_{\mathrm{e}p}$
and the coupling of nuclei to each other via $\mathbb{K}_{pq}$; in
particular, the screening of the nuclei by the inner electrons means
the tensors $\mathbb{K}_{pq}$ are typically only nonzero for nearest
neighbors~\cite{TianPRB2014, TianPRB2012, RuanPRB2006, CareyNMTE2008},
and also justifies our assumption that the nuclear oscillators do not
couple directly to long-range EM fields. All of these quantities
except the damping coefficients $b_{\mathrm{e}p}$ and
$b_{\mathrm{I}p}$ incorporate short-range interaction effects by
virtue of being computed from ab-initio DFT calculations or elemental
data; in principle, the damping coefficients may also be rigorously
derived by coupling these oscillator degrees of freedom to appropriate
reservoirs describing the full system, whether the molecule is in
vacuum, suspended in a condensed phase, or large enough to support a
continuum of phonons that irreversibly carry energy away, but we
choose simple approximate values justified by empirical considerations
appropriate to each system. These quantities are collected from all
molecules and respectively arranged into $3N \times 3N$ matrices
$(M_{\mathrm{e}}, Q_{\mathrm{e}}, B_{\mathrm{e}}, K_{\mathrm{e}},
M_{\mathrm{I}}, B_{\mathrm{I}}, K_{\mathrm{I}})$ that satisfy the
frequency-domain equations of motion~\cite{VenkataramPRL2017}
\begin{multline}
\begin{bmatrix}
  K_{\mathrm{e}} - \im\omega B_{\mathrm{e}} - \omega^{2} M_{\mathrm{e}} & -K_{\mathrm{e}} \\
  -K_{\mathrm{e}} & K_{\mathrm{e}} + K_{\mathrm{I}} - \im\omega
  B_{\mathrm{I}} - \omega^{2} M_{\mathrm{I}}
\end{bmatrix}
\begin{bmatrix}
  x_{\mathrm{e}} \\
  x_{\mathrm{I}}
\end{bmatrix}
\\ \hspace{-2in}=
\begin{bmatrix}
  Q_{\mathrm{e}} e_{\mathrm{e}} \\
  0
\end{bmatrix}
\end{multline}
for the electronic and nuclear oscillator displacements
$(x_{\mathrm{e}}, x_{\mathrm{I}})$ in terms of the total electric
field $\ket{\vec{E}}$ represented in the basis of electronic
oscillators as a $3N$-dimensional vector $e_{\mathrm{e}}$. Solving for
the electronic oscillator dipole moment
$p_{\mathrm{e}} = Q_{\mathrm{e}} x_{\mathrm{e}} = \alpha
e_{\mathrm{e}}$ yields the electric susceptibility matrix evaluated at
frequency $\omega = \im\xi$,
\begin{equation} \label{eq:alpha}
  \begin{split}
    \alpha &= Q_{\mathrm{e}} (K_{\mathrm{e}} + \xi B_{\mathrm{e}} + \xi^{2} M_{\mathrm{e}} \\ &- K_{\mathrm{e}} (K_{\mathrm{e}} + K_{\mathrm{I}} + \xi B_{\mathrm{I}} + \xi^{2} M_{\mathrm{I}})^{-1} K_{\mathrm{e}})^{-1} Q_{\mathrm{e}}
  \end{split}
\end{equation}
entering the expansion of $\mathbb{V}$ above. The nuclear and
electronic masses differ by four orders of magnitude, while their
respective harmonic coupling strengths are comparable, leading to two
frequency scales relevant to the response, with phonons dominating the
static and infrared response while valence electrons dominate at
visible and ultraviolet imaginary frequencies.

As mentioned above, even though we use DFT to account for the change
in electronic polarizability in each atom in a molecule due to its
neighbors, our use of oscillators to describe the valence electronic
degrees of freedom restricts our consideration to molecular systems
that are insulating or weakly metallic; at the level of the bare
susceptibility $\mathbb{V}$, the valence electronic oscillators by
themselves do not display the significant electronic delocalization
and mobility inherent in strongly metallic or doped systems, as may be
captured in tight-binding and related models. However, short-range
internuclear couplings encoded in $K_{\mathrm{I}}$ give rise to
collective nuclear oscillations (phonons) that in turn couple
different electronic oscillators to each other; this produces
nonlocality (spatial dispersion) in the bare susceptibility
$\mathbb{V}$ in an ab-initio rather than phenomenological way, and in
turn ensures that vdW interactions among atoms and molecules remain
finite even at vanishing separations, in contrast to point dipolar and
continuum treatments~\cite{BuhmannPRA2012, RodriguezNATURE2011,
  ZouNATURE2013}. In particular, we use Gaussian basis
functions~\cite{DonchevJCP2006, AmbrosettiSCIENCE2016,
  DiStasioJPCM2014, PhanJAP2013, ShtogunJPCL2010, KimLANGMUIR2007,
  ColeMS2009}
\begin{equation}
  \vec{f}_{pi} (\vec{x}) = \left(\sqrt{2\pi} \sigma_{p}\right)^{-3} \mathrm{exp}\left[-\frac{(\vec{x} - \vec{r}_{p})^{2}}{2\sigma_{p}^{2}}\right] \vec{e}_{i}
\end{equation}
centered at the equilibrium atomic positions $\vec{r}_{p}$, polarized
along the Cartesian direction $\vec{e}_{i}$, and normalized so that
$|\int \vec{f}_{pi} (\vec{x})~\mathrm{d}^{3} x| = 1$. The isotropic
widths $\sigma_{p}$ of these basis functions have been chosen at each
frequency as,
\begin{equation} \label{eq:sigmap} \sigma_{p} (\im\xi) =
  \frac{1}{\sqrt{4\pi}}\left(\frac{|\alpha_{p}(\im\xi)|}{3}\right)^{1/3},
\end{equation}
such that a dipolar oscillator of isotropic polarizability
$\alpha_{p}$ has the same self-interaction energy
$-\frac{\xi^{2}}{3c^{2}} \sum_i \bracket{\vec{f}_{pi}|\mathbb{G}_{0}
  \vec{f}_{pi}}$ as that of a Gaussian dipole distribution in
vacuum. Without phonons, the bare molecular response $\mathbb{V}$
would be essentially local, and each valence electronic oscillator
would have an individual polarizability
$\alpha_{p} (\im\xi) = q_{\mathrm{e}p}^{2} / (k_{\mathrm{e}p} + \xi
b_{\mathrm{e}p} + \xi^{2} m_{\mathrm{e}p})$, producing widths
$\sigma \sim 1$ angstrom at low frequency. With phonons, $\mathbb{V}$
is inherently nonlocal, so we extend~\eqref{sigmap} by defining
effective local isotropic atomic polarizabilities
\begin{equation}
  \alpha_{p} (\im\xi) = \frac{1}{3}\sum_{q,j} \alpha_{pj,qj} (\im\xi)
\end{equation}
that capture the spatial extent of nonlocal response due to phonons in
the atomistic systems we consider, especially at lower frequencies,
directly from the properties of the electronic and nuclear
oscillators; this atomic contraction effectively constitutes a local
approximation to the susceptibility within our oscillator
model~\cite{HermannCR2017}, but used only for constructing the
Gaussian widths. The resulting Gaussian widths strongly depend on the
molecular geometry, as encoded in nearest-neighbor bonds in the
internuclear coupling matrix $K_{\mathrm{I}}$. For low-dimensional
materials like carbyne or graphene, we observe $\sigma \sim 1$~nm at
low frequencies, with smaller widths arising in compact molecules like
fullerenes. 

Electromagnetic interactions among electronic oscillators are modified
by phonons in two ways, both of which depend strongly on the atomistic
geometry and material properties encoded in the oscillator model. The
first is that the basis functions $\ket{\vec{f}_{pi}}$ attenuate
short-distance EM divergences via the smearing of the valence
electronic response. Namely, EM fields can be significantly screened
if phonons enhance the magnitude and nonlocality of the electronic
response such that $\sigma$ increases beyond a few bond lengths. The
second arises from multiple scattering and screening effects through
the mutual coupling of the oscillators via long-range EM
fields~\cite{TkatchenkoJCP2013, DiStasioJPCM2014,
  AmbrosettiSCIENCE2016, HermannCR2017}; this gives rise to effective
plasmon--polaritons, as seen through shifts in the oscillator
frequencies from the poles of the resulting nonlocal susceptibility,
\begin{equation} \label{eq:chiposition}
  \tensor{\chi}(i\xi,\vec{x},\vec{x}')= -\frac{c^{2}}{\xi^{2}}
  \bracket{\vec{x} | \mathbb{V} | \vec{x}'} = \sum_{p,i,q,j}
  \alpha_{pi,qj} \vec{f}_{pi}(\vec{x}) \otimes \vec{f}_{qj}(\vec{x}'),
\end{equation}
which includes
only short-range interactions, to those of $\mathbb{T}$, which include
long-range EM interactions as well. This effect is present even in the
absence of phonons, in which case the nuclei are taken to be fixed in
space as in past work. With phonons, nonlocality may interplay in
complex ways with molecular geometry and material properties,
producing phonon--polariton resonances that couple nuclear,
electronic, and EM degrees of freedom, further modifying the poles of
$\mathbb{T}$. As we show below, the screening of long-range EM
interactions by phonons can modify the behavior of vdW interactions at
different distance regimes, especially as a function of
temperature. The inherent nonlocality in $\mathbb{V}$ due to
$K_{\mathrm{I}}$ also implies that in contrast to previous treatments
based on local electronic polarizabilities, the Born approximation of
the integrand $\Phi$ in~\eqref{vdWenergy} to low order in EM scattering
no longer represents a pairwise summation (PWS) approximation, as the
former includes correlations among all atoms due to phonons.

In the following, we study the RMB free energy of interaction at zero
and room temperatures between various molecular systems and a planar
gold slab, the latter of which we model as a perfect conductor for
simplicity in computing $\mathbb{G}_{\mathrm{env}}$ (via image
theory). Specifically, we consider (separately) a
$\mathrm{C}_{500}$-fullerene of radius $1$~nm, a long carbyne wire
oriented parallel to the surface, and an undoped graphene sheet of
infinite extent parallel to the surface; we focus on carbon allotropes
because their static electronic polarizabilities are large enough for
many-body effects to become particularly evident compared to
insulating biological molecules, but their electron densities are
localized enough for Hirshfeld partitioning and our oscillator model
to remain accurate compared to stronger metals. In each case, we make
appropriate comparisons to other predictions which typically make
further simplifications beyond our stated assumptions. We compare
results both in the presence and absence of phonons, the latter of
which is obtained by replacing $\alpha$ with a purely electronic
response
$\alpha_{\mathrm{e}} = Q_{\mathrm{e}} (K_{\mathrm{e}} + \xi
B_{\mathrm{e}} + \xi^{2} M_{\mathrm{e}})^{-1} Q_{\mathrm{e}}$, which
modifies the corresponding basis functions, $\mathbb{T}$, and
$\mathbb{T}_{\infty}$, and leads to a free energy which we denote
$\mathcal{F}_{\mathrm{e}}$. In addition, we also compare with the
Casimir--Polder (CP) approximate free energy
$\mathcal{F}_{\mathrm{CP}}$, which replaces~\eqref{vdWenergy} with,
\begin{equation}
  \mathcal{F}_{\mathrm{CP}} = -k_{\mathrm{B}}T \sum_{n = 0}^{\infty} {}'
  \trace{\tensor{\alpha}^{\mathrm{CP}} (\im\xi_{n}) \cdot
    \tensor{G}_{\mathrm{sca}} (\im\xi_{n}, \vec{r}, \vec{r})}
\label{eq:CP}
\end{equation}
in terms of
$\mathbb{G}_{\mathrm{sca}} = \mathbb{G}_{\mathrm{env}} -
\mathbb{G}_{0}$, representing a contraction of the molecule to a
dipole at position $\vec{r}$ with an effective polarizability
$\alpha^{\mathrm{CP}}_{ij} = \sum_{p,q}
\bracket{\vec{f}_{pi}|\mathbb{T}_{l\infty} \vec{f}_{qj}}$ that
includes long-range interactions among molecular degrees of freedom in
vacuum to infinite order in the scattering, but only to lowest order
with the surface. Finally, in the case of graphene, we compare to
predictions obtained via common macroscopic (continuum)
models~\cite{WunschNJP2006, WengerPRB2016, DobsonSS2011, HwangPRB2007,
  LundebergSCIENCE2017, SerneliusPRB2012}, described in detail in the
appendix.

\begin{figure}[t!]
\centering
\includegraphics[width=0.9\columnwidth]{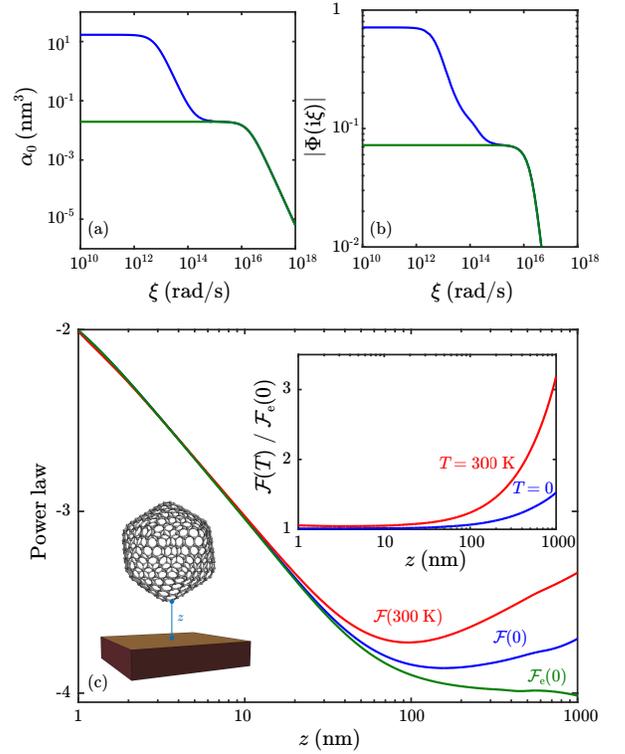}
\caption{\textbf{Impact of phonons at large separations.}  (a)
  Representative polarizability of an individual atomic constituent of
  a fullerene molecule suspended above a gold plate by a
  surface--surface gap $z$, comparing the full (including phonons)
  $\alpha$ (blue) and purely electronic $\alpha_{\mathrm{e}}$ (green)
  polarizabilities. (b) RMB free energy integrand $\Phi(i\xi)$ as a
  function of imaginary frequency $\xi$ corresponding to $\mathcal{F}$
  (blue) and $\mathcal{F}_{\mathrm{e}}$ (green) at a fixed
  $z = 1$~nm. (c) RMB power laws for $\mathcal{F}(0)$ (blue),
  $\mathcal{F}(300~\mathrm{K})$ (red), and
  $\mathcal{F}_{\mathrm{e}}(0)$ (green). Inset: energy ratios
  $\mathcal{F}(T) / \mathcal{F}_{\mathrm{e}}$ for the fullerene, at
  $T = 0$ (blue) or $T = 300~\mathrm{K}$ (red).}
\label{fig:C500fullerene}
\end{figure}

\begin{figure*}[t!]
\centering
\includegraphics[width=1\textwidth]{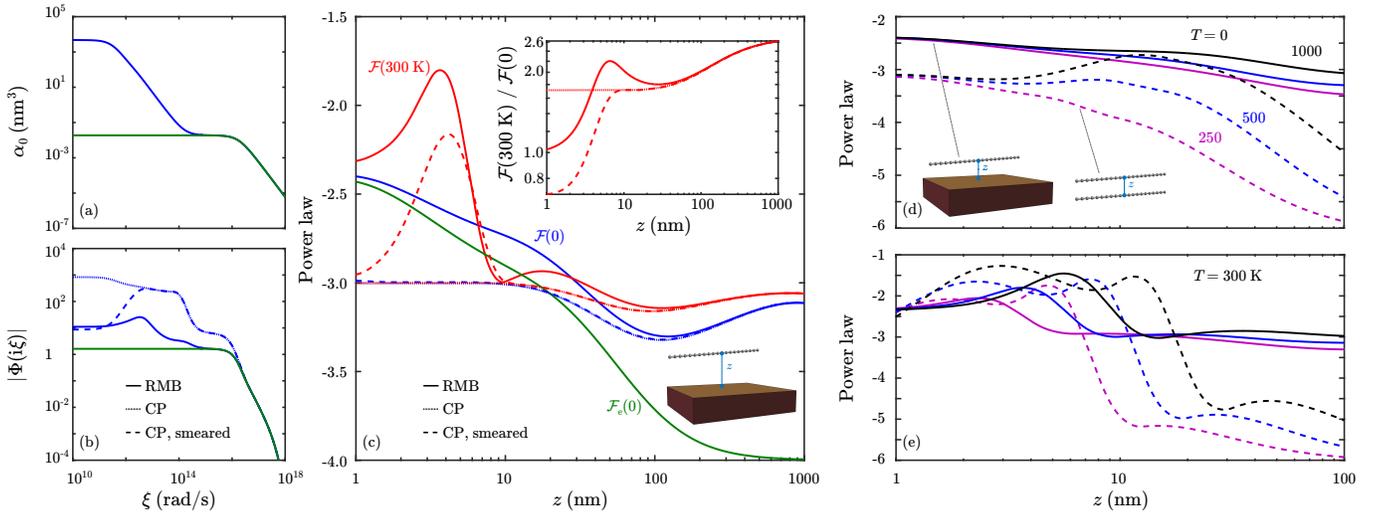}
\caption{\textbf{Nonmonotonicity and temperature deviations due to
    phonon-induced nonlocal response in an elongated molecule}. (a)
  Polarizability as a function of imaginary frequency for the middle
  atom (0) in a 500 atom-long carbyne wire, comparing $\alpha$ (blue)
  to $\alpha_{\mathrm{e}}$ (green). (b) Imaginary frequency integrands
  for $\mathcal{F}$ (blue) and $\mathcal{F}_{\mathrm{e}}$ (green) at
  $z = 1$~nm via RMB (solid) or CP, without (fine dashed) or
  with (coarse dashed) artificial smearing. (c) RMB (solid) and CP,
  without (fine dashed) or with (coarse dashed) artificial smearing,
  interaction power laws of a 500 atom-long carbyne wire parallel to a
  gold plate, for $\mathcal{F}(0)$ (blue),
  $\mathcal{F}(300~\mathrm{K})$ (red), and
  $\mathcal{F}_{\mathrm{e}}(0)$ (green). Inset: Free energy ratios
  $\mathcal{F}(300~\mathrm{K})/\mathcal{F}(0)$ as functions of $z$ via
  RMB (solid) or CP, without (fine dashed) or with (coarse dashed)
  artificial smearing. (d, e) Power laws for the vdW interactions of
  one parallel carbyne wire with a gold plate (solid) or two such
  wires in vacuum (dashed) for wires made of 250 (magenta), 500
  (blue), or 1000 (black) atoms, at $T = 0$ (d) or
  $T = 300~\mathrm{K}$ (e).}
\label{fig:carbyne}
\end{figure*}

\textbf{Fullerene}.---We begin with the case of a ``0-dimensional''
fullerene above the gold surface~[\figref{C500fullerene}]. An isolated
fullerene will not have the same dissipation mechanisms as a fullerene
in solution~\cite{BuhmannPRA2012}, so we neglect the damping
coefficients $B_{\mathrm{e}}$ and $B_{\mathrm{I}}$; however, we
constrain the center of mass of this isolated molecule by fixing the
positions of two nuclei on opposite sides of the fullerene. At large
$z$, the finite size of the fullerene is negligible, so it may be
treated like a point dipole with respect to scattering from the
plate. Without phonons, $\alpha_{\mathrm{e}}$ is characterized by a
single frequency scale arising from the electronic
response~[\figref{C500fullerene}(a)], so when the cutoff frequency
$c/z$ falls below that as $z$ increases~[\figref{C500fullerene}(b)],
the power law monotonically approaches the retarded dipolar limit of
$-4$ at $T = 0$. At room temperature, $T = 300~\mathrm{K}$, the power
law~[\figref{C500fullerene}(c)] will eventually increase to $-3$ only
when $z$ becomes comparable to the thermal wavelength,
$\hbar c/(k_{\mathrm{B}} T) \approx 7.6~\mathrm{\mu m}$, because once
the cutoff frequency $c/z$ falls below the first Matsubara frequency
$2\pi k_{\mathrm{B}} T/\hbar$, only the zero Matsubara frequency will
contribute. These predictions are similar to predictions from
macroscopic formulations of Casimir physics~\cite{BuhmannPRA2012};
namely, in the absence of phonons, the free energies at $T = 0$ versus
$T = 300~\mathrm{K}$ are essentially identical for
$z \leq 1~\mathrm{\mu m}$, like in typical macroscopic situations.

Matters change drastically when phonons are considered, in which case
$\alpha$ is characterized by two frequency scales due to the vastly
different nuclear and electronic masses. Even for
$z \leq 1~\mathrm{\mu m}$, as $z$ increases, these frequency scales
compete with the cutoff frequency $c/z$ to produce pronounced
nonmonotonic interaction power laws for $T = 0$ and
$T = 300~\mathrm{K}$; the onset of this deviation based on temperature
occurs at separations far smaller than one would expect from common
macroscopic predictions, though at large $z > 7.6~\mathrm{\mu m}$
(which we do not show) the asymptotic power laws approach those
observed in the absence of phonons. Such significant sensitivity to
temperature at small $z \approx 100$~nm illustrates that even
for a small, compact molecule like fullerene, the interplay between
phononic response and long-range EM fields can make the interaction
power laws deviate significantly from typical macroscopic
predictions. The increased relative importance of phononic response at
large $z$ also leads to strong deviations of the free energy ratios
$\mathcal{F}(T)/\mathcal{F}_{\mathrm{e}}(0)$ from 1 for large $z$ at
both values of $T$~[\figref{C500fullerene}(c)]. At smaller
separations, the finite size and curved spherical shape of the
fullerene dominate the interaction power law. Due to the small size of
the fullerene, at larger frequencies $\sim c/z$, phonons neither
significantly delocalize the molecular response nor strongly couple to
EM fields, consistent with the fact that Gaussian widths throughout
the molecule are smaller than the smallest value of $z=1$~nm
considered here. Hence, all three power laws converge upon each other
for $z \in [1~\mathrm{nm}, 10~\mathrm{nm}]$, and the free energy
ratios converge to 1 in this limit too. It is worth noting that at
$z \approx 100$~nm, where the power laws exhibit nonmonotonic
behavior, the vertical vdW force on the fullerene by the surface is on
the order of $10^{-18}~\mathrm{N}$, which is far smaller than
currently measurable in state-of-the-art vdW or Casimir force
experiments~\cite{WagnerNATURE2014, TsoiACSNANO2014, TangNATURE2017,
  GarrettPRL2018}.

\textbf{Carbyne}.---We now turn to the case of a ``1-dimensional''
carbyne wire~[\figref{carbyne}]. As with the fullerene, we assume the
damping coefficients $B_{\mathrm{e}}$ and $B_{\mathrm{I}}$ to be
negligible when the molecule is in isolation, and constrain the nuclei
at each end of the wire to remain fixed. While the qualitative
behaviors of the interaction power laws
$\partial \ln(\mathcal{F})/\partial \ln(z)$ for a wire at large $z$
above a gold plate are similar to those of the fullerene, the
elongated shape of the wire allows it to support longer-wavelength
phonons which couple much more strongly to low-frequency and infrared
EM fields, leading to a richer dependence on separation and
temperature even for $z < 100$~nm. To better understand the
physics of these interactions, we compare the RMB predictions to those
from the CP approximation of \eqref{CP}, first for a 500 atom-long
carbyne wire above a gold plate, and then for various configurations
of wire lengths (both in vacuum and above a plate).

For a 500 atom-long carbyne wire, delocalization in the response due
to phonons has the strongest effect at low frequencies, leading to
static Gaussian basis function widths
$\sigma_{0}(\omega\to 0) \approx 3.3$~nm at the middle of the
wire. \Figref{carbyne}(a) plots $\alpha_{0} \sim \sigma_0(\im\xi)^3$
as a function of $\xi$, showing the existence of two characteristic
frequency scales arising from the much stronger phonon-induced
electronic delocalization at infrared wavelengths. In contrast, the
response in the absence of phonons exhibits only a single frequency
scale and the Gaussian widths never exceed 1 angstrom. These enlarged
Gaussian smearing widths lead to nonmonotonicity in the RMB
integrand~[\figref{carbyne}(b)] with respect to $\xi$ for
$z < \sigma_{0}(0)$, as the Gaussian basis functions overlap with the
response of the gold plate (i.e. interactions with image basis
functions in the perfectly conducting limit). Such non-monotonicity
cannot be observed in the absence of phonons, as the bare
susceptibility $\tensor{\chi}(i\xi,\vec{x})$ is essentially local in
that case. However, as this nonmonotonicity occurs for very small
$\xi$ in the integrand, the behavior of the RMB power law at $T = 0$
with phonons with respect to $z$ is less sensitive to the nonmonotonic
integrand, simply approaching the power law without phonons as $z$
decreases~[\figref{carbyne}(c)]. By contrast, at $T = 300~\mathrm{K}$,
the RMB power law with phonons shows significant deviations from that
at $T = 0$ even up to $z < 20$~nm; essentially, the sampling of
Matsubara frequencies at room temperature makes the free energy
disproportionately sensitive to the response at static and infrared
$\xi$. Additionally, the sensitivity of these vdW interactions at room
temperature to the response at infrared and smaller $\xi$ leads to a
free energy ratio $\mathcal{F}(T)/\mathcal{F}(0)$ that exceeds 2 even
for $z < 20$~nm, and that energy ratio is nonmonotonic because the
zero and room temperature RMB free energy power laws with phonons
cross each other. 
At large $z$, the magnitude of the RMB free energy ratio
$\mathcal{F}(300~\mathrm{K})/\mathcal{F}(0)$ is consistent with recent
work by Maghrebi \emph{et al}.~\cite{MaghrebiPNAS2011}, which shows
the sensitivity of this energy ratio to geometry for continuum objects
even in the absence of nonlocal response. Note that at $z = 4$~nm, the
vertical force on the wire at $T = 0$ is
$F_{z} \approx 10^{-11}~\mathrm{N}$, and as can be observed from the
power laws and energy ratios, the force ratio is
$\frac{F_{z} (300~\mathrm{K})}{F_{z} (0)} \approx 1.2$; both the
forces themselves and their differences with respect to temperature
should therefore be measurable and resolvable in state-of-the-art
Casimir experiments~\cite{SerneliusJPA2006}, though it should be
pointed out that that long free-standing carbyne wires have not been
stably fabricable, and carbyne has only been found in solution or
confined to supramolecular structures like carbon
nanotubes~\cite{ChalifouxNATURE2010, ShiNATURE2016}.

To better understand these phenomena, we compare these results to
those obtained by the CP approximation, where the phonon polaritonic
response is contracted into a point dipolar polarizability. In the CP
approximation, the integrand is always monotonic, and the power laws
at each temperature are monotonic for $z < 20$~nm,
approaching the nonretarded dipolar limit of -3 as $z$ decreases. This
is because in the point dipolar limit, even if the magnitude of the
polarizability is enhanced due to phonon polaritons, there is no sense
in which nonlocality is captured in its long-range EM
interactions. The monotonic decrease in the CP integrand over a very
small frequency range also means that even at small $z$, Matsubara
summation at $T = 300~\mathrm{K}$ leads to a significantly larger
energy due to contributions from small $\xi$ than continuous
integration over $\xi$ at $T = 0$, so the energy ratio is
significantly larger than 1. However, if one artificially smears this
point dipole into an isotropic Gaussian distribution of width
$\sigma_{0}(\im\xi)$, such that the dipole at small $\xi$ will overlap
with its image in the conducting plane for $z < \sigma_{0}(0)$, one
qualitatively recovers the nonmonotonic integrand and room temperature
power laws. Notably, however, such a ``smeared CP approximation''
still leads to quantitative differences compared to RMB, as it
neglects explicit consideration of the finite molecular size and
geometry. For $z > \sigma_{0}(0)$, the smeared CP approximation
produces power laws identical to those of the standard CP
approximation, and both converge to the corresponding RMB power laws
at each respective temperature at much larger $z$.

Strictly speaking, for a wire parallel to a conducting plane, modeling
the latter via a local macroscopic susceptibility becomes questionable
for $z < \sigma_{0}(0)$, as we expect the atomism and spatially
dispersive response of the latter to matter more for such small
separations. Such an issue is not relevant when considering
interactions between two molecules in vacuum. We further explore this
by comparing the RMB interaction power laws (in the presence of
phonons) of a single carbyne wire above the gold plate, equivalent to
a wire interacting with its correlated image, against that of two
parallel, uncorrelated wires interacting in vacuum. In particular, we
study wires comprising of either 250, 500, or 1000 atoms at zero
~[\figref{carbyne}(d)] and room~[\figref{carbyne}(e)] temperatures. At
$T = 0$, the power laws for a single wire above the plate are all
monotonic even at small $z$, because the free energy is not sensitive
to the nonmonotonic integrand for $z < \sigma_{0}(0)$; essentially,
the wire is interacting with its correlated image, which dramatically
changes the phonon polaritons emerging from the long-range EM
interactions, compared to those of the wire in vacuum. In contrast,
the power laws for two wires of at least 500 atoms in vacuum show
nonmonotonicity for $z > 10$~nm, which is larger than
$\sigma_{0}(0)$ and hence cannot be attributed to overlapping Gaussian
basis functions. At $T = 300~\mathrm{K}$, the power laws for a single
wire above the plate show nonmonotonicity only for
$z \lesssim \sigma_{0}(0)$ for every wire length, while the power laws
for two wires in vacuum show two maxima for $z < 20$~nm, with
the one at larger $z$ corresponding to the aforementioned maxima
visible for two wires even at $T = 0$ and occurring in the absence of
overlapping Gaussian widths. Thus, it is clear that as nonmonotonic
vdW interaction power laws can be observed at room temperature for
separations both on the order or larger than the corresponding
Gaussian smearing widths, and is therefore not an artifact of
overlapping response functions or the lack of atomism in our
description of the plate.

\begin{figure}[t!]
\centering
\includegraphics[width=0.9\columnwidth]{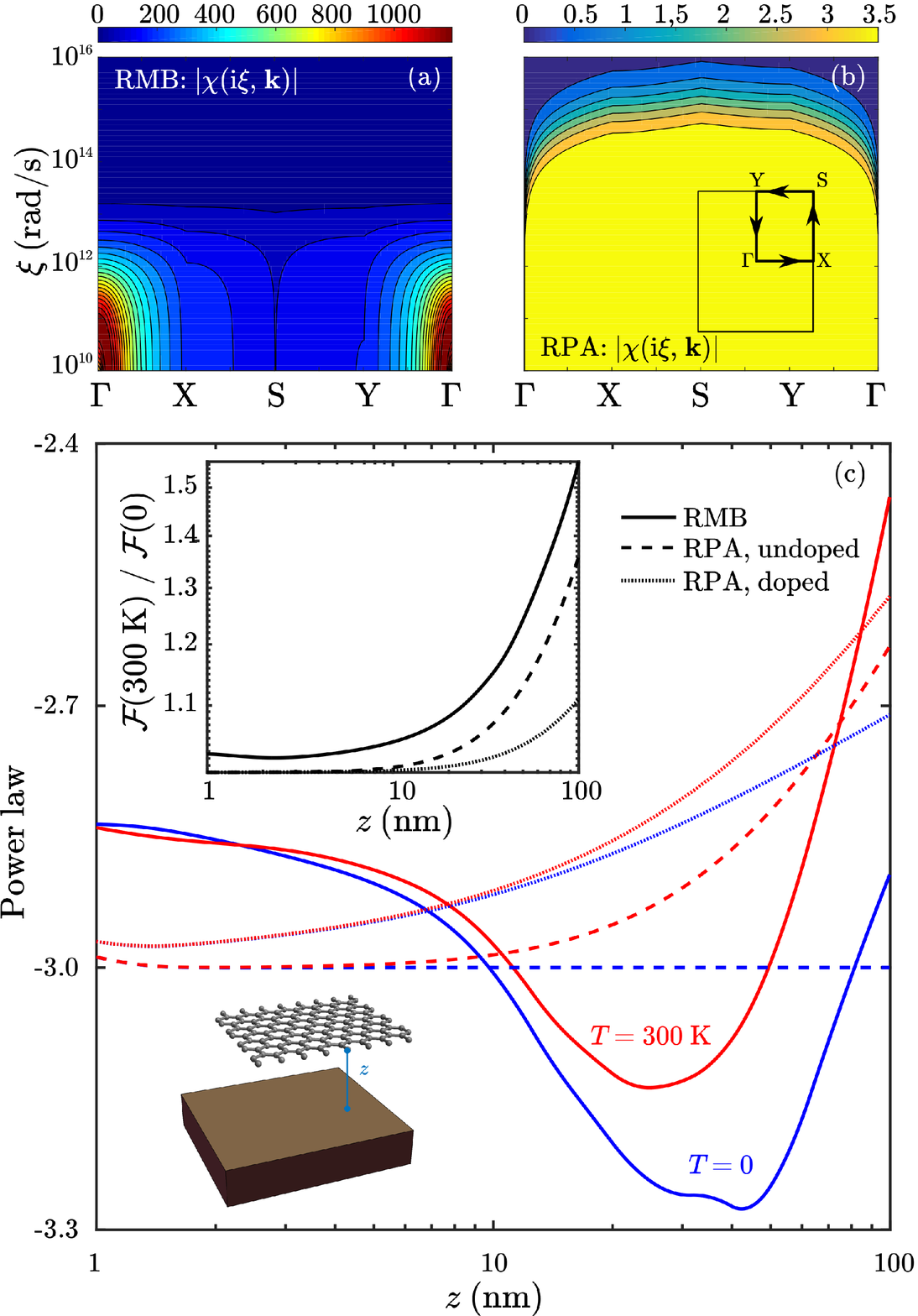}
\caption{\textbf{Impact of nonlocality and phonons on graphene vdW
    interactions.}  Magnitude of the Fourier-space susceptibility
  $|\chi(\im\xi, \vec{k})|$ of a pure (undoped) graphene sheet with
  rectangular unit cell 3.9~nm~$\times$~3.4~nm, obtained via either
  (a) RMB or (b) macroscopic, random-phase approximation
  (RPA)~\cite{SerneliusPRB2012} models. (c) Power law of interaction
  free energy for a graphene sheet suspended above gold plate by a
  vacuum gap $z$, at $T = 0$ (blue) or $T = 300~\mathrm{K}$ (red),
  comparing RMB (solid) results to macroscopic
  RPA
  predictions, either with (fine dashed) or without (coarse dashed)
  doping, where the doping concentration
  $n = 10^{13}~\mathrm{cm}^{-2}$. The inset shows the interaction free
  energy ratios $\mathcal{F}(300~\mathrm{K}) / \mathcal{F}(0)$ as a
  function of $z$.}
\label{fig:graphene}
\end{figure}

\textbf{Graphene}.---Finally, we consider a ``2-dimensional'' infinite
graphene sheet above the gold plate. Unlike fullerene molecules or
carbyne wires, atomically thin graphene sheets can be exfoliated and
suspended in vacuum~\cite{YiJMCA2015, BunchSCIENCE2007}; therefore,
compared to the other aforementioned carbon allotropes, there is
significantly more theoretical and experimental work characterizing
the mechanical and vibrational~\cite{FalkovskyJETP2007,
  EndlichPRB2013, AkinwandeEML2017},
electronic~\cite{CastroNetoRMP2009}, and
thermal~\cite{BalandinNATURE2011} properties of graphene. In the
particular context of vdW interactions, several theoretical
macroscopic models for the response of graphene have been employed,
which we summarily describe in the appendix. In what follows, we
compare our predictions to those obtained from a Lifshitz formula of
the Casimir interaction between a graphene and a plate, based on a
tight-binding model of the electronic band structure of graphene. Such
a model is consistent with the random phase approximation
(RPA)~\cite{SerneliusPRB2012} and includes spatial dispersion and the
possibility of doping, but does not consider contributions from
phonons or dissipation.

As with fullerene and carbyne, the RMB model of the response of
graphene partitions the quantum ground-state density into atomic
fragments, which are then mapped to a set of effective valence
electronic and nuclear oscillators. While the ground-state density
encodes similar physics contained in tight-binding
Hamiltonians~\cite{WunschNJP2006, WengerPRB2016, DobsonSS2011}, the
RMB framework require consideration of long-range EM interactions to
observe the emergence of electron delocalization (plasmon
polaritons). Moreover, the use of localized electron densities in the
oscillator model restricts us to consideration of undoped graphene
sheets, in contrast to the RPA model~\cite{SerneliusPRB2012}.
Meanwhile, our explicit consideration of phonons as well as
dissipation stands in contrast to the model used in
Ref.~\cite{SerneliusPRB2012}. While we do not explicitly show this,
our anisotropic internuclear couplings $K_{\mathrm{I}}$ derived from
DFT produce phonon dispersion relations that agree with prior
theoretical and experimental works~\cite{FalkovskyJETP2007,
  EndlichPRB2013}. Additionally, compared to the fullerene or carbyne
wire, we expect graphene to have more channels for dissipation in its
response. However, in undoped graphene, the dissipation rates due to
electron--electron or electron--acoustic phonon scattering typically
do not exceed $10^{12}~\mathrm{rad/s}$~\cite{LiPRB2018}. For
computational convenience, we employ a somewhat larger dissipation
rate of $10^{13}~\mathrm{rad/s}$, encoded in $B_{\mathrm{I}}$. Such
large dissipation rates have been considered before in the context of
strongly (electronically) doped graphene and observed in measurements
of DC mobility~\cite{IlicPRB2012}. In any case, our use of larger
damping rate is justifiable if the graphene sample is very impure,
even if it is not electronically doped.

We begin by demonstrating the differences in the graphene nonlocal
susceptibility (ignoring long-range EM interactions) predicted at
imaginary frequency in our RMB framework~[\figref{graphene}(a)]
compared to the RPA model~[\figref{graphene}(b)] in the absence of
doping. A Fourier space representation of $\tensor{\chi}$ may be
obtained from its position space representation,
$ \tensor{\chi}(\im\xi, \vec{k}, \vec{k}') = \int
\mathrm{d}^3\vec{x}~\mathrm{d}^3\vec{x}'~ \tensor{\chi}(\im\xi,
\vec{x}, \vec{x}') e^{-\im(\vec{k} \cdot \vec{x} - \vec{k}' \cdot
  \vec{x}')}$. As discussed in~\appref{RPAresponse}, the RPA model
assumes a continuum electronic susceptibility derived from the
long-wavelength, conical electronic band structure of graphene in a
tight-binding approximation. The resulting susceptibility is isotropic
and diagonal in Fourier space,
$\tensor{\chi}(\im\xi, \vec{k}, \vec{k}') = (2\pi)^{3} \chi(\im\xi,
\vec{k}) \tensor{1} \delta(\vec{k} - \vec{k}')$, or equivalently,
$\chi(\im\xi, \vec{k}) = \frac{1}{3} \int
\mathrm{d}^3\vec{k}'/(2\pi)^{3}~ \trace{\tensor{\chi}(\im\xi, \vec{k},
  \vec{k}')}$, where $\chi(\im\xi, \vec{k})$ is both real and
positive. By contrast, the atomism of graphene is explicitly accounted
for in RMB via our choice of atomic basis functions, in which case
$\tensor{\chi}(\im\xi, \vec{k}, \vec{k}')$ is not diagonal. In
particular, applying the definitions above to the RMB
model~\eqref{chiposition}, yields the following definition for the
complex RMB susceptibility:
\begin{multline}
  \chi(\im\xi, \vec{k}) = \frac{1}{3} \sum_{p,i,q,j}
  \alpha_{pi,qi}(\im\xi) \left(\int \mathrm{d}^{3}\vec{x}~ \vec{f}_{pi}
    (\vec{x}) e^{-\im\vec{k} \cdot \vec{x}} \right) \cdot \vec{f}_{qj}
  (0)
\end{multline}
Given the atomistic nature of the nonlocality in RMB, there is some
ambiguity in our choice of the coordinate $\vec{x}'=0$, which we
choose to be the center of a hexagonal honeycomb cell. Note also that
we compute the RMB response of an infinite sheet of graphene by
applying Bloch's theorem to determine the ground state electron
density and associated quantities within a 3.9~nm~$\times$~3.4~nm
rectangular unit cell, in which case the quantity of interest is the
free energy per unit area.

In the RMB model, at small $\xi$, and especially at small $\vec{k}$,
corresponding to nearly spatially uniform incident fields,
contributions from long-wavelength acoustic phonons tend to dominate,
delocalizing the electronic response over several primitive unit cells
and drastically increasing $|\chi|$ to over 100; as $|\vec{k}|$
increases, so too do the contributions of localized optical phonons,
decreasing the magnitude and nonlocality of the response. It is only
for $\xi > 10^{14}~\mathrm{rad/s}$ that phonons no longer contribute
appreciably to the response, so $|\chi|$ is essentially independent of
$\vec{k}$ and its magnitude is less than 10, consistent with RPA and
decaying quickly as $\xi$ increases further. In contrast to RMB, the
electronic orbitals in the RPA model are tightly bound to fixed
nuclei, in which case the resulting response vanishes as
$\vec{k} \to 0$ for nonzero frequency, and hence $\chi$ increases with
increasing $|\vec{k}|$, in qualitative opposition to the RMB
prediction. In particular, in the short-wavelength regime, $\chi$
approaches a constant asymptote
$\frac{q_{\mathrm{e}}^{2} g}{32\hbar\epsilon_{0} v_{\mathrm{F}}}
\approx 4$ at a rate controlled by $\xi$, that is far smaller than the
RMB prediction at low $\xi$. Given that the $\pi$-orbitals within RPA
are tightly bound to nearby nuclei, one would expect that phonons
should have a significant impact upon the valence electronic and
static response of graphene. The low-frequency behavior of metals is
particularly important in the context of vdW interactions, as
exemplified by recent discrepancies in the response of Drude versus
plasma models of gold~\cite{SerneliusJPA2006}. While our discussion of
the RPA model has thus far neglected doping, if doping is considered
even to an arbitrarily small degree, the presence of free mobile
charge carriers (plasmons) does lead to a divergence in the RPA
susceptibility as $\vec{k} \to 0$, providing slightly better
qualitative agreement with RMB predictions.

We now compare the predictions of both RMB and RPA models for the vdW
interaction free energy per unit area of graphene above a gold plate,
at both zero and room temperatures. In particular,
\figref{graphene}(c) shows the free energy power law and corresponding
energy ratios, with respect to the gap separation $z$. The RPA model
is analyzed in the presence and absence of doping, with doping
concentration $n = 10^{13}~\mathrm{cm}^{-2}$. Our RMB predictions for
the power laws are qualitatively similar to those of the fullerene in
that the delocalization in the response due to phonons leads to power
laws at both temperatures that are nonmonotonic, though the higher
dimensionality of graphene compared to fullerene or carbyne causes the
onset of nonmonotonic behavior to arise at smaller $z \gtrsim 20$~nm
than for the fullerene or carbyne; while issues of convergence and
numerical error limit our consideration to $z \leq 100$~nm, we expect
the RMB power laws to asymptotically approach -3 and -2 at zero and
finite temperature, respectively. Likewise, as the impact of phonons
is more pronounced at larger temperatures, the energy ratio starts to
deviate noticeably from 1 at $z \gtrsim 20$~nm. Note that at
$z = 100$~nm, the vertical force ratio
$F_{z} (300~\mathrm{K})/F_{z} (0) \approx 1.3$ and corresponding
pressure $F_{z}/A \approx 1~\mathrm{N/m^{2}}$, and thus these
differences should be measurable in state-of-the-art
experiments~\cite{BanishevPRB2013, KlimchitskayaPRB2015}. Compared to
our previous predictions in carbyne systems, the assumption of much
larger nuclear dissipation $B_{\mathrm{I}}$ in graphene significantly
dampens and smears the impact of long-wavelength acoustic phonons,
limiting the phonon mean free path and hence the spatial extent of
delocalization in our Gaussian widths $\sigma$ to a much greater
degree. Consequently, we observe energy ratios much closer to 1 and
monotonic power laws at small separations $z \lesssim 20$~nm. Notably,
the RPA power law without doping is a constant -3 at zero temperature
over all $z$, increasing monotonically toward -2 at finite
temperatures. With doping, both power laws are observed to increase
from -3 for $z > 10$~nm, though as Sernelius~\cite{SerneliusPRB2012}
makes clear, these begin to approach their asymptotic values at very
large $z \gg 100$~nm. Finally, since RPA does not include phonons,
there is no nonmonotonic behavior in the power laws, and the RPA power
laws with or without doping are consistently above -3 in the range of
separations $z$ that we consider, unlike the RMB power laws which drop
below -3 for $z \sim 20$~nm.

\textbf{Concluding remarks}.---We have demonstrated the strong
influence of nonlocal response arising from phonons on the vdW
interactions of molecular systems, particularly highlighting the
dependence of these effects on molecular shape, size, temperature, and
material dissipation. These effects can conspire to produce changes in
the vdW interaction energy, relative to treatments that neglect
phonons or finite temperature, which should be measurable with
state-of-the-art experiments, particularly at nanometric separations
where macroscopic treatments of Casimir forces in bulk
media~\cite{SerneliusJPA2006, SushkovNATURE2011} predict insignificant
temperature effects. The characteristics of molecular vibrations
(phonons) in our calculations were derived from covalent bond
properties, so one might expect more delocalization of electronic
response along the bonds than perpendicular to them. This implies that
further accuracy in modeling could be achieved by using anisotropic
Gaussian widths to account for this anisotropy. For example, in
carbyne, the ratio of transverse to longitudinal internuclear spring
coupling coefficients between nearest neighbors is 0.04, while this
difference is approximately 0.35 in graphene, so the strong anisotropy
in the nonlocal response should be captured in the Gaussian widths in
order to more accurately model the material response at short
separations. Finally, a more accurate comparison between the RMB
predictions and those of macroscopic models for undoped graphene
requires smaller dissipation rates, which is likely to result in more
temperature-sensitive energies (potentially leading to non-monotonic
behavior at small $z$ akin to those observed in carbyne
wires). However, the computational complexity of simulating large unit
cells supporting more strongly delocalized phonons makes such a
demonstration challenging. Furthermore, intuitive interpretation of
the RPA power laws is hampered by the complicated form of the
susceptibility, while comparisons with RMB for doped graphene would
require a reformulation of the localized oscillator model to allow for
inherently delocalized electronic response along with associated
changes to DFT computations, the subject of future work.

\emph{Acknowledgements}.---This work was supported by the National
Science Foundation under Grants No. DMR-1454836, DMR 1420541, DGE
1148900, and the Cornell Center for Materials Research MRSEC (award
no. DMR-1719875), as well as the Luxembourg National Research within
the FNR-CORE program (No. FNR-11360857). PSV thanks Chinmay Khandekar,
Weiliang Jin, and Sean Molesky for helpful discussions.

\appendix
\section{Model of molecular response} \label{sec:RMBresponse}

The ground-state (i.e. zero-temperature) electron density and
geometric configuration of nuclei are determined for each molecule $l$
separately in isolation by minimizing the energy of the molecule
computed via density functional theory (DFT) in the Born--Oppenheimer
approximation: for each set of fixed nuclear coordinates, the
ground-state electron density is obtained via DFT, and through this,
the nuclear coordinates are varied to reach an overall minimum energy,
so this process produces ground-state densities and nuclear
coordinates that account for short-range quantum exchange,
correlation, hybridization, and electrostatic effects. In the
insulating or weakly metallic molecular systems that we consider, the
electronic wavefunctions are somewhat localized, allowing for a
Hirshfeld partitioning of the ground state electron density over the
molecule into atomic fragments that account for the presence of other
nearby atoms; these atomic fragments are then used with the electron
densities of the corresponding isolated atoms to produce static
electronic polarizabilities $\alpha_{\mathrm{e}p}(0)$ associated with
each atom $p$.

As illustrated schematically in~\figref{schematic}, for each molecule
$l$, we map onto the set of $N_{l}$ atoms labeled by $p$ (denoting the
total number of atoms by $N = \sum_{l} N_{l}$) a set of coupled
harmonic oscillator degrees of freedom. In particular, each atom $p$
consists of a single nuclear oscillator, representing the nucleus
screened by the inner electrons, and a single electronic oscillator
representing the effective valence electrons; this effectively
represents an Uns\"{o}ld approximation~\cite{HermannCR2017}, which in
principle could be relaxed by assigning multiple oscillators to
account for many different possible electronic transitions. Within the
Uns\"{o}ld approximation, for the purpose of constructing the
molecular response, the electronic oscillator in each atom is
initially taken to be undamped (with dissipation to be added later);
given $\alpha_{\mathrm{e}p}(0)$, the electronic oscillator frequencies
$\omega_{\mathrm{e}p}$ are computed by fitting the oscillator
dispersion to nonretarded vdW $C_{6}$-coefficients for each atom taken
from a large reference of theoretical and experimental atomic and
small molecular data~\cite{TkatchenkoJCP2013, DiStasioJPCM2014,
  HermannCR2017}. From this, the effective number of electrons
$n_{\mathrm{e}p}$ associated with that atom can be determined, and so
can the effective charge
$q_{\mathrm{e}p} = n_{\mathrm{e}p} q_{\mathrm{e}}$, mass
$m_{\mathrm{e}p} = n_{\mathrm{e}p} m_{\mathrm{e}}$, and isotropic
spring constant $k_{\mathrm{e}p}$; these quantities, by virtue of
deriving from $\alpha_{\mathrm{e}p}(0)$ and $\omega_{\mathrm{e}p}$,
encode the same short-range quantum and electrostatic effects present
in DFT and other high-level quantum calculations~\cite{DonchevJCP2006,
  AmbrosettiSCIENCE2016, DiStasioJPCM2014, PhanJAP2013,
  ShtogunJPCL2010, KimLANGMUIR2007, ColeMS2009}. The nuclear masses
are taken from elemental data as they are four orders of magnitude
larger than the electronic masses, while the internuclear spring
constants $\mathbb{K}_{pq}$ are computed as the second spatial
derivatives of the ground-state energy in DFT with respect to the
nuclear coordinates at equilibrium.

\section{Model of macroscopic graphene
  response} \label{sec:RPAresponse}

Macroscopic treatments of vdW or Casimir interactions involving
graphene rely on continuum models of its optical susceptibility, which
enter into the familiar Lifshitz formula as idealized reflection
coefficients of perfectly thin sheets~\cite{WoodsRMP2015}. These
models typically start with quantum-mechanical tight-binding
Hamiltonian for the localized $\pi$-bonding
orbitals~\cite{WunschNJP2006, WengerPRB2016, DobsonSS2011,
  HwangPRB2007, LundebergSCIENCE2017, SerneliusPRB2012}, while
neglecting contributions from phonons to the material response and
dissipation (though such contributions can be mitigated quantum
mechanically through the addition of appropriate
coupling~\cite{JablanPRB2011} and reservoir~\cite{LiPRB2018}
potentials.) This quantum-mechanical tight-binding electronic
Hamiltonian is also commonly approximated as having linear dispersion
around the Dirac points, in which case the susceptibility is derived
as the lowest-order linear response to an applied perturbative
electric field, consistent with the random-phase approximation
(RPA). While such a framework is typically presented at zero
temperature for analytical convenience, the addition of a Fermi-Dirac
distribution in the formula for the linear response allows for
consideration of finite temperature effects as well as doping (by
varying the chemical potential) on the bare response. Alternative
treatments start with the second-quantized Dirac Hamiltonian of
graphene electrons and fields near the Dirac points, ultimately
recovering similar expressions for the response in the presence or
absence of doping~\cite{BarlasPRL2007, BordagPRB2009,
  BordagPRB2016}. The resulting expressions for the linear EM response
explicitly show spatial dispersion. However, only models of doped
graphene (which are independent of p- versus n-type doping for the
same doping concentration) seem to explicitly consider dissipation, as
changing the doping concentration allows for more dissipative
mechanisms (e.g. electron-electron or electron-phonon scattering),
especially when the chemical potential is far from the Dirac point or
when finite temperature is directly used in the construction of the
bare response. In the context of vdW interactions, it is also common
to simplify these expressions by taking the limit of vanishing
parallel wavevector, thereby neglecting spatial dispersion and
resulting in relatively small nonlocal length scales in the frequency
ranges of interest. Consequently, the response functions follow a
similar form as that of Drude or plasma susceptibilities, and in the
particular case of undoped graphene, it is common to further
approximate the conductivity as having the universal constant
$q_{\mathrm{e}}^{2} / (16\pi\hbar\epsilon_{0})$ over a large range of
frequency scales~\cite{DrosdoffPRB2010, KoppensNANOLETT2011,
  IlicPRB2012}. vdW and Casimir interactions involve integrals over
all frequencies~\eqref{vdWenergy}, in which case the infrared response
(including both temporal and spatial dispersion) is expected to be
relevant.

To better understand vdW interactions between a graphene sheet and a
gold surface, we compare the susceptibility of graphene in our RMB
model to an appropriate macroscopic counterpart in the main text. In
particular, we use the RPA-derived response of
Sernelius~\cite{SerneliusPRB2012} (referred to in the main text simply
as RPA), with wavevector-dependent permittivity
$\epsilon(\omega, \vec{k}) = 1 + \chi(\omega, \vec{k})$ defined in
terms of the susceptibility
\begin{equation}
  \chi(\omega, \vec{k}) = \frac{q_{\mathrm{e}}^{2} g|\vec{k}|}{32\hbar\epsilon_{0} \sqrt{v_{\mathrm{F}}^{2} |\vec{k}|^{2} - \omega^{2}}}
\end{equation}
in the undoped case, or
\begin{equation}
  \begin{split}
    \chi(\omega, \vec{k}) &= \frac{q_{\mathrm{e}}^{2} D_{0}}{2\epsilon_{0} |\vec{k}|} \left(1 + \frac{\kappa^{2}}{4\sqrt{\kappa^{2} - \zeta^{2}}} (\pi - \phi(\kappa, \zeta))\right) \\
    \phi(\kappa, \zeta) &= \arcsin((1 - \zeta)/\kappa) + \arcsin((1 + \zeta)/\kappa) \\
    &- \frac{\zeta - 1}{\kappa} \sqrt{1 - \left(\frac{\zeta -
          1}{\kappa}\right)^{2}} + \frac{\zeta + 1}{\kappa} \sqrt{1 -
      \left(\frac{\zeta + 1}{\kappa}\right)^{2}}
  \end{split}
\end{equation}
in the doped case, for $\vec{k}$ in the 
plane. These are defined in terms of the Fermi velocity
$v_{\mathrm{F}} = 8.73723\times 10^{5}~\mathrm{m/s}$ and the
spin-pseudospin degeneracy $g = 4$, as well as the Fermi wavevector
$k_{\mathrm{F}} = \sqrt{4\pi |n|/g}$, Fermi energy
$E_{\mathrm{F}} = \hbar v_{\mathrm{F}} k_{\mathrm{F}}$, electron
density of states
$D_{0} = \frac{\sqrt{g|n|/\pi}}{\hbar v_{\mathrm{F}}}$ at the Fermi
level, and dimensionless variables
$\kappa = |\vec{k}|/(2k_{\mathrm{F}})$ and
$\zeta = \hbar\omega/(2E_{\mathrm{F}})$. We use these expressions as
they capture temporal and spatial dispersion in graphene, while being
simpler to manipulate than equivalent expressions derived from second
quantization~\cite{BordagPRB2009, BordagPRB2016}. These expressions
for the undoped and doped macroscopic response may be used for direct
comparison with the RMB susceptibility as well as for computing vdW
interactions as compared with corresponding RMB predictions; for the
latter, as mentioned above, the RPA susceptibilities are used to
construct reflection coefficients, which are in turn used in the
Lifshitz formula for the Casimir interaction between parallel planar
surfaces.

\nocite{apsrev41Control}
\bibliographystyle{apsrev4-1}

\bibliography{molvdwphononpaper2}

\begin{thebibliography}{69}%
\makeatletter
\providecommand \@ifxundefined [1]{%
 \@ifx{#1\undefined}
}%
\providecommand \@ifnum [1]{%
 \ifnum #1\expandafter \@firstoftwo
 \else \expandafter \@secondoftwo
 \fi
}%
\providecommand \@ifx [1]{%
 \ifx #1\expandafter \@firstoftwo
 \else \expandafter \@secondoftwo
 \fi
}%
\providecommand \natexlab [1]{#1}%
\providecommand \enquote  [1]{``#1''}%
\providecommand \bibnamefont  [1]{#1}%
\providecommand \bibfnamefont [1]{#1}%
\providecommand \citenamefont [1]{#1}%
\providecommand \href@noop [0]{\@secondoftwo}%
\providecommand \href [0]{\begingroup \@sanitize@url \@href}%
\providecommand \@href[1]{\@@startlink{#1}\@@href}%
\providecommand \@@href[1]{\endgroup#1\@@endlink}%
\providecommand \@sanitize@url [0]{\catcode `\\12\catcode `\$12\catcode
  `\&12\catcode `\#12\catcode `\^12\catcode `\_12\catcode `\%12\relax}%
\providecommand \@@startlink[1]{}%
\providecommand \@@endlink[0]{}%
\providecommand \url  [0]{\begingroup\@sanitize@url \@url }%
\providecommand \@url [1]{\endgroup\@href {#1}{\urlprefix }}%
\providecommand \urlprefix  [0]{URL }%
\providecommand \Eprint [0]{\href }%
\providecommand \doibase [0]{http://dx.doi.org/}%
\providecommand \selectlanguage [0]{\@gobble}%
\providecommand \bibinfo  [0]{\@secondoftwo}%
\providecommand \bibfield  [0]{\@secondoftwo}%
\providecommand \translation [1]{[#1]}%
\providecommand \BibitemOpen [0]{}%
\providecommand \bibitemStop [0]{}%
\providecommand \bibitemNoStop [0]{.\EOS\space}%
\providecommand \EOS [0]{\spacefactor3000\relax}%
\providecommand \BibitemShut  [1]{\csname bibitem#1\endcsname}%
\let\auto@bib@innerbib\@empty
\bibitem [{\citenamefont {Woods}\ \emph {et~al.}(2016)\citenamefont {Woods},
  \citenamefont {Dalvit}, \citenamefont {Tkatchenko}, \citenamefont
  {Rodriguez-Lopez}, \citenamefont {Rodriguez},\ and\ \citenamefont
  {Podgornik}}]{WoodsRMP2015}%
  \BibitemOpen
  \bibfield  {author} {\bibinfo {author} {\bibfnamefont {L.~M.}\ \bibnamefont
  {Woods}}, \bibinfo {author} {\bibfnamefont {D.~A.~R.}\ \bibnamefont
  {Dalvit}}, \bibinfo {author} {\bibfnamefont {A.}~\bibnamefont {Tkatchenko}},
  \bibinfo {author} {\bibfnamefont {P.}~\bibnamefont {Rodriguez-Lopez}},
  \bibinfo {author} {\bibfnamefont {A.~W.}\ \bibnamefont {Rodriguez}}, \ and\
  \bibinfo {author} {\bibfnamefont {R.}~\bibnamefont {Podgornik}},\ }\bibfield
  {title} {\enquote {\bibinfo {title} {Materials perspective on casimir and van
  der waals interactions},}\ }\href {\doibase 10.1103/RevModPhys.88.045003}
  {\bibfield  {journal} {\bibinfo  {journal} {Rev. Mod. Phys.}\ }\textbf
  {\bibinfo {volume} {88}},\ \bibinfo {pages} {045003} (\bibinfo {year}
  {2016})}\BibitemShut {NoStop}%
\bibitem [{\citenamefont {Langbein}(1974)}]{Langbein1974}%
  \BibitemOpen
  \bibfield  {author} {\bibinfo {author} {\bibfnamefont {D.}~\bibnamefont
  {Langbein}},\ }\enquote {\bibinfo {title} {Theory of van der waals
  attraction},}\ in\ \href {\doibase 10.1007/BFb0042407} {\emph {\bibinfo
  {booktitle} {Springer Tracts in Modern Physics}}}\ (\bibinfo  {publisher}
  {Springer Berlin Heidelberg},\ \bibinfo {address} {Berlin, Heidelberg},\
  \bibinfo {year} {1974})\ pp.\ \bibinfo {pages} {1--139}\BibitemShut {NoStop}%
\bibitem [{\citenamefont {Tkatchenko}(2015)}]{TkatchenkoADFM2014}%
  \BibitemOpen
  \bibfield  {author} {\bibinfo {author} {\bibfnamefont {A.}~\bibnamefont
  {Tkatchenko}},\ }\bibfield  {title} {\enquote {\bibinfo {title} {Current
  understanding of van der waals effects in realistic materials},}\ }\href
  {\doibase 10.1002/adaaaaafm.201403029} {\bibfield  {journal} {\bibinfo
  {journal} {Advanced Functional Materials}\ }\textbf {\bibinfo {volume}
  {25}},\ \bibinfo {pages} {2054--2061} (\bibinfo {year} {2015})}\BibitemShut
  {NoStop}%
\bibitem [{\citenamefont {Tkatchenko}\ \emph {et~al.}(2013)\citenamefont
  {Tkatchenko}, \citenamefont {Ambrosetti},\ and\ \citenamefont {{DiStasio
  Jr.}}}]{TkatchenkoJCP2013}%
  \BibitemOpen
  \bibfield  {author} {\bibinfo {author} {\bibfnamefont {A.}~\bibnamefont
  {Tkatchenko}}, \bibinfo {author} {\bibfnamefont {A.}~\bibnamefont
  {Ambrosetti}}, \ and\ \bibinfo {author} {\bibfnamefont {R.~A.}\ \bibnamefont
  {{DiStasio Jr.}}},\ }\bibfield  {title} {\enquote {\bibinfo {title}
  {Interatomic methods for the dispersion energy derived from the adiabatic
  connection fluctuation-dissipation theorem},}\ }\href {\doibase
  http://dx.doi.org/10.1063/1.4789814} {\bibfield  {journal} {\bibinfo
  {journal} {The Journal of Chemical Physics}\ }\textbf {\bibinfo {volume}
  {138}},\ \bibinfo {pages} {074106} (\bibinfo {year} {2013})}\BibitemShut
  {NoStop}%
\bibitem [{\citenamefont {Gobre}\ and\ \citenamefont
  {Tkatchenko}(2013)}]{GobreNCOMMS2013}%
  \BibitemOpen
  \bibfield  {author} {\bibinfo {author} {\bibfnamefont {V.~V.}\ \bibnamefont
  {Gobre}}\ and\ \bibinfo {author} {\bibfnamefont {A.}~\bibnamefont
  {Tkatchenko}},\ }\bibfield  {title} {\enquote {\bibinfo {title} {Scaling laws
  for van der waals interactions in nanostructured materials},}\ }\href
  {\doibase http://dx.doi.org/10.1038/ncomms3341} {\bibfield  {journal}
  {\bibinfo  {journal} {Nature Communications}\ }\textbf {\bibinfo {volume}
  {4}} (\bibinfo {year} {2013}),\
  http://dx.doi.org/10.1038/ncomms3341}\BibitemShut {NoStop}%
\bibitem [{\citenamefont {{DiStasio Jr.}}\ \emph {et~al.}(2014)\citenamefont
  {{DiStasio Jr.}}, \citenamefont {Gobre},\ and\ \citenamefont
  {Tkatchenko}}]{DiStasioJPCM2014}%
  \BibitemOpen
  \bibfield  {author} {\bibinfo {author} {\bibfnamefont {R.~A.}\ \bibnamefont
  {{DiStasio Jr.}}}, \bibinfo {author} {\bibfnamefont {V.~V.}\ \bibnamefont
  {Gobre}}, \ and\ \bibinfo {author} {\bibfnamefont {A.}~\bibnamefont
  {Tkatchenko}},\ }\bibfield  {title} {\enquote {\bibinfo {title} {Many-body
  van der waals interactions in molecules and condensed matter},}\ }\href
  {http://stacks.iop.org/0953-8984/26/i=21/a=213202} {\bibfield  {journal}
  {\bibinfo  {journal} {Journal of Physics: Condensed Matter}\ }\textbf
  {\bibinfo {volume} {26}},\ \bibinfo {pages} {213202} (\bibinfo {year}
  {2014})}\BibitemShut {NoStop}%
\bibitem [{\citenamefont {Ambrosetti}\ \emph {et~al.}(2016)\citenamefont
  {Ambrosetti}, \citenamefont {Ferri}, \citenamefont {DiStasio},\ and\
  \citenamefont {Tkatchenko}}]{AmbrosettiSCIENCE2016}%
  \BibitemOpen
  \bibfield  {author} {\bibinfo {author} {\bibfnamefont {A.}~\bibnamefont
  {Ambrosetti}}, \bibinfo {author} {\bibfnamefont {N.}~\bibnamefont {Ferri}},
  \bibinfo {author} {\bibfnamefont {R.~A.}\ \bibnamefont {DiStasio},
  \bibfnamefont {Jr.}}, \ and\ \bibinfo {author} {\bibfnamefont
  {A.}~\bibnamefont {Tkatchenko}},\ }\bibfield  {title} {\enquote {\bibinfo
  {title} {Wavelike charge density fluctuations and van der waals interactions
  at the nanoscale},}\ }\href {\doibase 10.1126/science.aae0509} {\bibfield
  {journal} {\bibinfo  {journal} {Science}\ }\textbf {\bibinfo {volume}
  {351}},\ \bibinfo {pages} {1171--1176} (\bibinfo {year} {2016})}\BibitemShut
  {NoStop}%
\bibitem [{\citenamefont {Reilly}\ and\ \citenamefont
  {Tkatchenko}(2014)}]{ReillyPRL2014}%
  \BibitemOpen
  \bibfield  {author} {\bibinfo {author} {\bibfnamefont {A.~M.}\ \bibnamefont
  {Reilly}}\ and\ \bibinfo {author} {\bibfnamefont {A.}~\bibnamefont
  {Tkatchenko}},\ }\bibfield  {title} {\enquote {\bibinfo {title} {Role of
  dispersion interactions in the polymorphism and entropic stabilization of the
  aspirin crystal},}\ }\href {\doibase 10.1103/PhysRevLett.113.055701}
  {\bibfield  {journal} {\bibinfo  {journal} {Phys. Rev. Lett.}\ }\textbf
  {\bibinfo {volume} {113}},\ \bibinfo {pages} {055701} (\bibinfo {year}
  {2014})}\BibitemShut {NoStop}%
\bibitem [{\citenamefont {Reilly}\ and\ \citenamefont
  {Tkatchenko}(2015)}]{ReillyCS2015}%
  \BibitemOpen
  \bibfield  {author} {\bibinfo {author} {\bibfnamefont {A.~M.}\ \bibnamefont
  {Reilly}}\ and\ \bibinfo {author} {\bibfnamefont {A.}~\bibnamefont
  {Tkatchenko}},\ }\bibfield  {title} {\enquote {\bibinfo {title} {van der
  waals dispersion interactions in molecular materials: beyond pairwise
  additivity},}\ }\href {\doibase 10.1039/C5SC00410A} {\bibfield  {journal}
  {\bibinfo  {journal} {Chem. Sci.}\ }\textbf {\bibinfo {volume} {6}},\
  \bibinfo {pages} {3289--3301} (\bibinfo {year} {2015})}\BibitemShut {NoStop}%
\bibitem [{\citenamefont {Hoja}\ \emph {et~al.}(2017)\citenamefont {Hoja},
  \citenamefont {Reilly},\ and\ \citenamefont {Tkatchenko}}]{HojaCMS2017}%
  \BibitemOpen
  \bibfield  {author} {\bibinfo {author} {\bibfnamefont {J.}~\bibnamefont
  {Hoja}}, \bibinfo {author} {\bibfnamefont {A.~M.}\ \bibnamefont {Reilly}}, \
  and\ \bibinfo {author} {\bibfnamefont {A.}~\bibnamefont {Tkatchenko}},\
  }\bibfield  {title} {\enquote {\bibinfo {title} {First-principles modeling of
  molecular crystals: structures and stabilities, temperature and pressure},}\
  }\href {\doibase 10.1002/wcms.1294} {\bibfield  {journal} {\bibinfo
  {journal} {Wiley Interdisciplinary Reviews: Computational Molecular Science}\
  }\textbf {\bibinfo {volume} {7}},\ \bibinfo {pages} {e1294--n/a} (\bibinfo
  {year} {2017})},\ \bibinfo {note} {e1294}\BibitemShut {NoStop}%
\bibitem [{\citenamefont {Buhmann}\ \emph {et~al.}(2012)\citenamefont
  {Buhmann}, \citenamefont {Scheel}, \citenamefont {Ellingsen}, \citenamefont
  {Hornberger},\ and\ \citenamefont {Jacob}}]{BuhmannPRA2012}%
  \BibitemOpen
  \bibfield  {author} {\bibinfo {author} {\bibfnamefont {S.~Y.}\ \bibnamefont
  {Buhmann}}, \bibinfo {author} {\bibfnamefont {S.}~\bibnamefont {Scheel}},
  \bibinfo {author} {\bibfnamefont {S.~A.}\ \bibnamefont {Ellingsen}}, \bibinfo
  {author} {\bibfnamefont {K.}~\bibnamefont {Hornberger}}, \ and\ \bibinfo
  {author} {\bibfnamefont {A.}~\bibnamefont {Jacob}},\ }\bibfield  {title}
  {\enquote {\bibinfo {title} {Casimir-polder interaction of fullerene
  molecules with surfaces},}\ }\href {\doibase 10.1103/PhysRevA.85.042513}
  {\bibfield  {journal} {\bibinfo  {journal} {Phys. Rev. A}\ }\textbf {\bibinfo
  {volume} {85}},\ \bibinfo {pages} {042513} (\bibinfo {year}
  {2012})}\BibitemShut {NoStop}%
\bibitem [{\citenamefont {Rodriguez}\ \emph {et~al.}(2011)\citenamefont
  {Rodriguez}, \citenamefont {Capasso},\ and\ \citenamefont
  {Johnson}}]{RodriguezNATURE2011}%
  \BibitemOpen
  \bibfield  {author} {\bibinfo {author} {\bibfnamefont {A.~W.}\ \bibnamefont
  {Rodriguez}}, \bibinfo {author} {\bibfnamefont {F.}~\bibnamefont {Capasso}},
  \ and\ \bibinfo {author} {\bibfnamefont {S.~G.}\ \bibnamefont {Johnson}},\
  }\bibfield  {title} {\enquote {\bibinfo {title} {The casimir effect in
  microstructured geometries},}\ }\href@noop {} {\bibfield  {journal} {\bibinfo
   {journal} {Nature Photonics}\ }\textbf {\bibinfo {volume} {5}},\ \bibinfo
  {pages} {211--221} (\bibinfo {year} {2011})}\BibitemShut {NoStop}%
\bibitem [{\citenamefont {Zou}\ \emph {et~al.}(2013)\citenamefont {Zou},
  \citenamefont {Marcet}, \citenamefont {Rodriguez}, \citenamefont {Reid},
  \citenamefont {McCauley}, \citenamefont {Kravchenko}, \citenamefont {Lu},
  \citenamefont {Bao}, \citenamefont {Johnson},\ and\ \citenamefont
  {Chan}}]{ZouNATURE2013}%
  \BibitemOpen
  \bibfield  {author} {\bibinfo {author} {\bibfnamefont {J.}~\bibnamefont
  {Zou}}, \bibinfo {author} {\bibfnamefont {Z.}~\bibnamefont {Marcet}},
  \bibinfo {author} {\bibfnamefont {A.~W.}\ \bibnamefont {Rodriguez}}, \bibinfo
  {author} {\bibfnamefont {M.~T.~H.}\ \bibnamefont {Reid}}, \bibinfo {author}
  {\bibfnamefont {A.~P.}\ \bibnamefont {McCauley}}, \bibinfo {author}
  {\bibfnamefont {I.~I.}\ \bibnamefont {Kravchenko}}, \bibinfo {author}
  {\bibfnamefont {T.}~\bibnamefont {Lu}}, \bibinfo {author} {\bibfnamefont
  {Y.}~\bibnamefont {Bao}}, \bibinfo {author} {\bibfnamefont {S.~G.}\
  \bibnamefont {Johnson}}, \ and\ \bibinfo {author} {\bibfnamefont {H.~B.}\
  \bibnamefont {Chan}},\ }\bibfield  {title} {\enquote {\bibinfo {title}
  {Casimir forces on a silicon micromechanical chip},}\ }\href@noop {}
  {\bibfield  {journal} {\bibinfo  {journal} {Nature communications}\ }\textbf
  {\bibinfo {volume} {4}},\ \bibinfo {pages} {1845} (\bibinfo {year}
  {2013})}\BibitemShut {NoStop}%
\bibitem [{\citenamefont {Dhar}\ and\ \citenamefont {Roy}(2006)}]{DharJSP2006}%
  \BibitemOpen
  \bibfield  {author} {\bibinfo {author} {\bibfnamefont {A.}~\bibnamefont
  {Dhar}}\ and\ \bibinfo {author} {\bibfnamefont {D.}~\bibnamefont {Roy}},\
  }\bibfield  {title} {\enquote {\bibinfo {title} {Heat transport in harmonic
  lattices},}\ }\href {\doibase 10.1007/s10955-006-9235-3} {\bibfield
  {journal} {\bibinfo  {journal} {Journal of Statistical Physics}\ }\textbf
  {\bibinfo {volume} {125}},\ \bibinfo {pages} {801--820} (\bibinfo {year}
  {2006})}\BibitemShut {NoStop}%
\bibitem [{\citenamefont {Tian}\ \emph {et~al.}(2014)\citenamefont {Tian},
  \citenamefont {Esfarjani},\ and\ \citenamefont {Chen}}]{TianPRB2014}%
  \BibitemOpen
  \bibfield  {author} {\bibinfo {author} {\bibfnamefont {Z.}~\bibnamefont
  {Tian}}, \bibinfo {author} {\bibfnamefont {K.}~\bibnamefont {Esfarjani}}, \
  and\ \bibinfo {author} {\bibfnamefont {G.}~\bibnamefont {Chen}},\ }\bibfield
  {title} {\enquote {\bibinfo {title} {Green's function studies of phonon
  transport across si/ge superlattices},}\ }\href {\doibase
  10.1103/PhysRevB.89.235307} {\bibfield  {journal} {\bibinfo  {journal} {Phys.
  Rev. B}\ }\textbf {\bibinfo {volume} {89}},\ \bibinfo {pages} {235307}
  (\bibinfo {year} {2014})}\BibitemShut {NoStop}%
\bibitem [{\citenamefont {Tian}\ \emph {et~al.}(2012)\citenamefont {Tian},
  \citenamefont {Esfarjani},\ and\ \citenamefont {Chen}}]{TianPRB2012}%
  \BibitemOpen
  \bibfield  {author} {\bibinfo {author} {\bibfnamefont {Z.}~\bibnamefont
  {Tian}}, \bibinfo {author} {\bibfnamefont {K.}~\bibnamefont {Esfarjani}}, \
  and\ \bibinfo {author} {\bibfnamefont {G.}~\bibnamefont {Chen}},\ }\bibfield
  {title} {\enquote {\bibinfo {title} {Enhancing phonon transmission across a
  si/ge interface by atomic roughness: First-principles study with the green's
  function method},}\ }\href {\doibase 10.1103/PhysRevB.86.235304} {\bibfield
  {journal} {\bibinfo  {journal} {Phys. Rev. B}\ }\textbf {\bibinfo {volume}
  {86}},\ \bibinfo {pages} {235304} (\bibinfo {year} {2012})}\BibitemShut
  {NoStop}%
\bibitem [{\citenamefont {Mingo}\ and\ \citenamefont
  {Yang}(2003)}]{MingoPRB2003}%
  \BibitemOpen
  \bibfield  {author} {\bibinfo {author} {\bibfnamefont {N.}~\bibnamefont
  {Mingo}}\ and\ \bibinfo {author} {\bibfnamefont {L.}~\bibnamefont {Yang}},\
  }\bibfield  {title} {\enquote {\bibinfo {title} {Phonon transport in
  nanowires coated with an amorphous material: An atomistic green's function
  approach},}\ }\href {\doibase 10.1103/PhysRevB.68.245406} {\bibfield
  {journal} {\bibinfo  {journal} {Phys. Rev. B}\ }\textbf {\bibinfo {volume}
  {68}},\ \bibinfo {pages} {245406} (\bibinfo {year} {2003})}\BibitemShut
  {NoStop}%
\bibitem [{\citenamefont {Chiloyan}\ \emph {et~al.}(2015)\citenamefont
  {Chiloyan}, \citenamefont {Garg}, \citenamefont {Esfarjani},\ and\
  \citenamefont {Chen}}]{ChiloyanNATURE2015}%
  \BibitemOpen
  \bibfield  {author} {\bibinfo {author} {\bibfnamefont {V.}~\bibnamefont
  {Chiloyan}}, \bibinfo {author} {\bibfnamefont {J.}~\bibnamefont {Garg}},
  \bibinfo {author} {\bibfnamefont {K.}~\bibnamefont {Esfarjani}}, \ and\
  \bibinfo {author} {\bibfnamefont {G.}~\bibnamefont {Chen}},\ }\bibfield
  {title} {\enquote {\bibinfo {title} {Transition from near-field thermal
  radiation to phonon heat conduction at sub-nanometre gaps},}\ }\href@noop {}
  {\bibfield  {journal} {\bibinfo  {journal} {Nature communications}\ }\textbf
  {\bibinfo {volume} {6}},\ \bibinfo {pages} {6755} (\bibinfo {year}
  {2015})}\BibitemShut {NoStop}%
\bibitem [{\citenamefont {Pendry}\ \emph {et~al.}(2016)\citenamefont {Pendry},
  \citenamefont {Sasihithlu},\ and\ \citenamefont {Craster}}]{PendryPRB2016}%
  \BibitemOpen
  \bibfield  {author} {\bibinfo {author} {\bibfnamefont {J.~B.}\ \bibnamefont
  {Pendry}}, \bibinfo {author} {\bibfnamefont {K.}~\bibnamefont {Sasihithlu}},
  \ and\ \bibinfo {author} {\bibfnamefont {R.~V.}\ \bibnamefont {Craster}},\
  }\bibfield  {title} {\enquote {\bibinfo {title} {Phonon-assisted heat
  transfer between vacuum-separated surfaces},}\ }\href {\doibase
  10.1103/PhysRevB.94.075414} {\bibfield  {journal} {\bibinfo  {journal} {Phys.
  Rev. B}\ }\textbf {\bibinfo {volume} {94}},\ \bibinfo {pages} {075414}
  (\bibinfo {year} {2016})}\BibitemShut {NoStop}%
\bibitem [{\citenamefont {Henry}\ and\ \citenamefont
  {Chen}(2008)}]{HenryPRL2008}%
  \BibitemOpen
  \bibfield  {author} {\bibinfo {author} {\bibfnamefont {A.}~\bibnamefont
  {Henry}}\ and\ \bibinfo {author} {\bibfnamefont {G.}~\bibnamefont {Chen}},\
  }\bibfield  {title} {\enquote {\bibinfo {title} {High thermal conductivity of
  single polyethylene chains using molecular dynamics simulations},}\ }\href
  {\doibase 10.1103/PhysRevLett.101.235502} {\bibfield  {journal} {\bibinfo
  {journal} {Phys. Rev. Lett.}\ }\textbf {\bibinfo {volume} {101}},\ \bibinfo
  {pages} {235502} (\bibinfo {year} {2008})}\BibitemShut {NoStop}%
\bibitem [{\citenamefont {Esfarjani}\ \emph {et~al.}(2011)\citenamefont
  {Esfarjani}, \citenamefont {Chen},\ and\ \citenamefont
  {Stokes}}]{EsfarjaniPRB2011}%
  \BibitemOpen
  \bibfield  {author} {\bibinfo {author} {\bibfnamefont {K.}~\bibnamefont
  {Esfarjani}}, \bibinfo {author} {\bibfnamefont {G.}~\bibnamefont {Chen}}, \
  and\ \bibinfo {author} {\bibfnamefont {H.~T.}\ \bibnamefont {Stokes}},\
  }\bibfield  {title} {\enquote {\bibinfo {title} {Heat transport in silicon
  from first-principles calculations},}\ }\href {\doibase
  10.1103/PhysRevB.84.085204} {\bibfield  {journal} {\bibinfo  {journal} {Phys.
  Rev. B}\ }\textbf {\bibinfo {volume} {84}},\ \bibinfo {pages} {085204}
  (\bibinfo {year} {2011})}\BibitemShut {NoStop}%
\bibitem [{\citenamefont {Gonz\'alez~Noya}\ \emph {et~al.}(2004)\citenamefont
  {Gonz\'alez~Noya}, \citenamefont {Srivastava}, \citenamefont
  {Chernozatonskii},\ and\ \citenamefont {Menon}}]{NoyaPRB2004}%
  \BibitemOpen
  \bibfield  {author} {\bibinfo {author} {\bibfnamefont {E.}~\bibnamefont
  {Gonz\'alez~Noya}}, \bibinfo {author} {\bibfnamefont {D.}~\bibnamefont
  {Srivastava}}, \bibinfo {author} {\bibfnamefont {L.~A.}\ \bibnamefont
  {Chernozatonskii}}, \ and\ \bibinfo {author} {\bibfnamefont {M.}~\bibnamefont
  {Menon}},\ }\bibfield  {title} {\enquote {\bibinfo {title} {Thermal
  conductivity of carbon nanotube peapods},}\ }\href {\doibase
  10.1103/PhysRevB.70.115416} {\bibfield  {journal} {\bibinfo  {journal} {Phys.
  Rev. B}\ }\textbf {\bibinfo {volume} {70}},\ \bibinfo {pages} {115416}
  (\bibinfo {year} {2004})}\BibitemShut {NoStop}%
\bibitem [{\citenamefont {Cui}\ \emph {et~al.}(2015)\citenamefont {Cui},
  \citenamefont {Feng},\ and\ \citenamefont {Zhang}}]{CuiJPCA2015}%
  \BibitemOpen
  \bibfield  {author} {\bibinfo {author} {\bibfnamefont {L.}~\bibnamefont
  {Cui}}, \bibinfo {author} {\bibfnamefont {Y.}~\bibnamefont {Feng}}, \ and\
  \bibinfo {author} {\bibfnamefont {X.}~\bibnamefont {Zhang}},\ }\bibfield
  {title} {\enquote {\bibinfo {title} {Dependence of thermal conductivity of
  carbon nanopeapods on filling ratios of fullerene molecules},}\ }\href
  {\doibase 10.1021/acs.jpca.5b07995} {\bibfield  {journal} {\bibinfo
  {journal} {The Journal of Physical Chemistry A}\ }\textbf {\bibinfo {volume}
  {119}},\ \bibinfo {pages} {11226--11232} (\bibinfo {year}
  {2015})}\BibitemShut {NoStop}%
\bibitem [{\citenamefont {Venkataram}\ \emph {et~al.}(2017)\citenamefont
  {Venkataram}, \citenamefont {Hermann}, \citenamefont {Tkatchenko},\ and\
  \citenamefont {Rodriguez}}]{VenkataramPRL2017}%
  \BibitemOpen
  \bibfield  {author} {\bibinfo {author} {\bibfnamefont {P.~S.}\ \bibnamefont
  {Venkataram}}, \bibinfo {author} {\bibfnamefont {J.}~\bibnamefont {Hermann}},
  \bibinfo {author} {\bibfnamefont {A.}~\bibnamefont {Tkatchenko}}, \ and\
  \bibinfo {author} {\bibfnamefont {A.~W.}\ \bibnamefont {Rodriguez}},\
  }\bibfield  {title} {\enquote {\bibinfo {title} {Unifying microscopic and
  continuum treatments of van der waals and casimir interactions},}\ }\href
  {\doibase 10.1103/PhysRevLett.118.266802} {\bibfield  {journal} {\bibinfo
  {journal} {Phys. Rev. Lett.}\ }\textbf {\bibinfo {volume} {118}},\ \bibinfo
  {pages} {266802} (\bibinfo {year} {2017})}\BibitemShut {NoStop}%
\bibitem [{\citenamefont {Sushkov}\ \emph {et~al.}(2011)\citenamefont
  {Sushkov}, \citenamefont {Kim}, \citenamefont {Dalvit},\ and\ \citenamefont
  {Lamoreaux}}]{SushkovNATURE2011}%
  \BibitemOpen
  \bibfield  {author} {\bibinfo {author} {\bibfnamefont {A.}~\bibnamefont
  {Sushkov}}, \bibinfo {author} {\bibfnamefont {W.}~\bibnamefont {Kim}},
  \bibinfo {author} {\bibfnamefont {D.}~\bibnamefont {Dalvit}}, \ and\ \bibinfo
  {author} {\bibfnamefont {S.}~\bibnamefont {Lamoreaux}},\ }\bibfield  {title}
  {\enquote {\bibinfo {title} {Observation of the thermal casimir force},}\
  }\href@noop {} {\bibfield  {journal} {\bibinfo  {journal} {Nature Physics}\
  }\textbf {\bibinfo {volume} {7}},\ \bibinfo {pages} {230} (\bibinfo {year}
  {2011})}\BibitemShut {NoStop}%
\bibitem [{\citenamefont {Wagner}\ \emph {et~al.}(2014)\citenamefont {Wagner},
  \citenamefont {Fournier}, \citenamefont {Ruiz}, \citenamefont {Li},
  \citenamefont {M{\"u}llen}, \citenamefont {Rohlfing}, \citenamefont
  {Tkatchenko}, \citenamefont {Temirov},\ and\ \citenamefont
  {Tautz}}]{WagnerNATURE2014}%
  \BibitemOpen
  \bibfield  {author} {\bibinfo {author} {\bibfnamefont {C.}~\bibnamefont
  {Wagner}}, \bibinfo {author} {\bibfnamefont {N.}~\bibnamefont {Fournier}},
  \bibinfo {author} {\bibfnamefont {V.~G.}\ \bibnamefont {Ruiz}}, \bibinfo
  {author} {\bibfnamefont {C.}~\bibnamefont {Li}}, \bibinfo {author}
  {\bibfnamefont {K.}~\bibnamefont {M{\"u}llen}}, \bibinfo {author}
  {\bibfnamefont {M.}~\bibnamefont {Rohlfing}}, \bibinfo {author}
  {\bibfnamefont {A.}~\bibnamefont {Tkatchenko}}, \bibinfo {author}
  {\bibfnamefont {R.}~\bibnamefont {Temirov}}, \ and\ \bibinfo {author}
  {\bibfnamefont {F.~S.}\ \bibnamefont {Tautz}},\ }\bibfield  {title} {\enquote
  {\bibinfo {title} {Non-additivity of molecule-surface van der waals
  potentials from force measurements},}\ }\href@noop {} {\bibfield  {journal}
  {\bibinfo  {journal} {Nature communications}\ }\textbf {\bibinfo {volume}
  {5}},\ \bibinfo {pages} {5568} (\bibinfo {year} {2014})}\BibitemShut
  {NoStop}%
\bibitem [{\citenamefont {Rahi}\ \emph {et~al.}(2009)\citenamefont {Rahi},
  \citenamefont {Emig}, \citenamefont {Graham}, \citenamefont {Jaffe},\ and\
  \citenamefont {Kardar}}]{RahiPRD2009}%
  \BibitemOpen
  \bibfield  {author} {\bibinfo {author} {\bibfnamefont {S.~J.}\ \bibnamefont
  {Rahi}}, \bibinfo {author} {\bibfnamefont {T.}~\bibnamefont {Emig}}, \bibinfo
  {author} {\bibfnamefont {N.}~\bibnamefont {Graham}}, \bibinfo {author}
  {\bibfnamefont {R.~L.}\ \bibnamefont {Jaffe}}, \ and\ \bibinfo {author}
  {\bibfnamefont {M.}~\bibnamefont {Kardar}},\ }\bibfield  {title} {\enquote
  {\bibinfo {title} {Scattering theory approach to electrodynamic casimir
  forces},}\ }\href {\doibase 10.1103/PhysRevD.80.085021} {\bibfield  {journal}
  {\bibinfo  {journal} {Phys. Rev. D}\ }\textbf {\bibinfo {volume} {80}},\
  \bibinfo {pages} {085021} (\bibinfo {year} {2009})}\BibitemShut {NoStop}%
\bibitem [{\citenamefont {Reid}\ \emph {et~al.}(2013)\citenamefont {Reid},
  \citenamefont {White},\ and\ \citenamefont {Johnson}}]{ReidPRA2013}%
  \BibitemOpen
  \bibfield  {author} {\bibinfo {author} {\bibfnamefont {M.~T.~H.}\
  \bibnamefont {Reid}}, \bibinfo {author} {\bibfnamefont {J.}~\bibnamefont
  {White}}, \ and\ \bibinfo {author} {\bibfnamefont {S.~G.}\ \bibnamefont
  {Johnson}},\ }\bibfield  {title} {\enquote {\bibinfo {title} {Fluctuating
  surface currents: An algorithm for efficient prediction of casimir
  interactions among arbitrary materials in arbitrary geometries},}\
  }\href@noop {} {\bibfield  {journal} {\bibinfo  {journal} {Phys. Rev. A}\
  }\textbf {\bibinfo {volume} {88}},\ \bibinfo {pages} {022514} (\bibinfo
  {year} {2013})}\BibitemShut {NoStop}%
\bibitem [{\citenamefont {Johnson}(2011)}]{Johnson2011}%
  \BibitemOpen
  \bibfield  {author} {\bibinfo {author} {\bibfnamefont {S.~G.}\ \bibnamefont
  {Johnson}},\ }\enquote {\bibinfo {title} {Numerical methods for computing
  casimir interactions},}\ in\ \href {\doibase 10.1007/978-3-642-20288-9_6}
  {\emph {\bibinfo {booktitle} {Casimir Physics}}},\ \bibinfo {editor} {edited
  by\ \bibinfo {editor} {\bibfnamefont {D.}~\bibnamefont {Dalvit}}, \bibinfo
  {editor} {\bibfnamefont {P.}~\bibnamefont {Milonni}}, \bibinfo {editor}
  {\bibfnamefont {D.}~\bibnamefont {Roberts}}, \ and\ \bibinfo {editor}
  {\bibfnamefont {F.}~\bibnamefont {da~Rosa}}}\ (\bibinfo  {publisher}
  {Springer Berlin Heidelberg},\ \bibinfo {address} {Berlin, Heidelberg},\
  \bibinfo {year} {2011})\ pp.\ \bibinfo {pages} {175--218}\BibitemShut
  {NoStop}%
\bibitem [{\citenamefont {Rodriguez}\ \emph {et~al.}(2015)\citenamefont
  {Rodriguez}, \citenamefont {Hui}, \citenamefont {Woolf}, \citenamefont
  {Johnson}, \citenamefont {Lon\v{c}ar},\ and\ \citenamefont
  {Capasso}}]{RodriguezADP2015}%
  \BibitemOpen
  \bibfield  {author} {\bibinfo {author} {\bibfnamefont {A.~W.}\ \bibnamefont
  {Rodriguez}}, \bibinfo {author} {\bibfnamefont {P.-C.}\ \bibnamefont {Hui}},
  \bibinfo {author} {\bibfnamefont {D.~P.}\ \bibnamefont {Woolf}}, \bibinfo
  {author} {\bibfnamefont {S.~G.}\ \bibnamefont {Johnson}}, \bibinfo {author}
  {\bibfnamefont {M.}~\bibnamefont {Lon\v{c}ar}}, \ and\ \bibinfo {author}
  {\bibfnamefont {F.}~\bibnamefont {Capasso}},\ }\bibfield  {title} {\enquote
  {\bibinfo {title} {Classical and fluctuation-induced electromagnetic
  interactions in micron-scale systems: designer bonding, antibonding, and
  casimir forces},}\ }\href@noop {} {\bibfield  {journal} {\bibinfo  {journal}
  {Annalen der Physik}\ }\textbf {\bibinfo {volume} {527}},\ \bibinfo {pages}
  {45--80} (\bibinfo {year} {2015})}\BibitemShut {NoStop}%
\bibitem [{\citenamefont {Ambrosetti}\ \emph {et~al.}(2014)\citenamefont
  {Ambrosetti}, \citenamefont {Reilly}, \citenamefont {DiStasio},\ and\
  \citenamefont {Tkatchenko}}]{AmbrosettiJCP2014}%
  \BibitemOpen
  \bibfield  {author} {\bibinfo {author} {\bibfnamefont {A.}~\bibnamefont
  {Ambrosetti}}, \bibinfo {author} {\bibfnamefont {A.~M.}\ \bibnamefont
  {Reilly}}, \bibinfo {author} {\bibfnamefont {R.~A.}\ \bibnamefont
  {DiStasio}}, \ and\ \bibinfo {author} {\bibfnamefont {A.}~\bibnamefont
  {Tkatchenko}},\ }\bibfield  {title} {\enquote {\bibinfo {title} {Long-range
  correlation energy calculated from coupled atomic response functions},}\
  }\href {\doibase http://dx.doi.org/10.1063/1.4865104} {\bibfield  {journal}
  {\bibinfo  {journal} {The Journal of Chemical Physics}\ }\textbf {\bibinfo
  {volume} {140}},\ \bibinfo {pages} {18A508} (\bibinfo {year}
  {2014})}\BibitemShut {NoStop}%
\bibitem [{\citenamefont {Ruan}\ and\ \citenamefont
  {Kaviany}(2006)}]{RuanPRB2006}%
  \BibitemOpen
  \bibfield  {author} {\bibinfo {author} {\bibfnamefont {X.~L.}\ \bibnamefont
  {Ruan}}\ and\ \bibinfo {author} {\bibfnamefont {M.}~\bibnamefont {Kaviany}},\
  }\bibfield  {title} {\enquote {\bibinfo {title} {Enhanced laser cooling of
  rare-earth-ion-doped nanocrystalline powders},}\ }\href {\doibase
  10.1103/PhysRevB.73.155422} {\bibfield  {journal} {\bibinfo  {journal} {Phys.
  Rev. B}\ }\textbf {\bibinfo {volume} {73}},\ \bibinfo {pages} {155422}
  (\bibinfo {year} {2006})}\BibitemShut {NoStop}%
\bibitem [{\citenamefont {Carey}\ \emph {et~al.}(2008)\citenamefont {Carey},
  \citenamefont {Chen}, \citenamefont {Grigoropoulos}, \citenamefont
  {Kaviany},\ and\ \citenamefont {Majumdar}}]{CareyNMTE2008}%
  \BibitemOpen
  \bibfield  {author} {\bibinfo {author} {\bibfnamefont {V.~P.}\ \bibnamefont
  {Carey}}, \bibinfo {author} {\bibfnamefont {G.}~\bibnamefont {Chen}},
  \bibinfo {author} {\bibfnamefont {C.}~\bibnamefont {Grigoropoulos}}, \bibinfo
  {author} {\bibfnamefont {M.}~\bibnamefont {Kaviany}}, \ and\ \bibinfo
  {author} {\bibfnamefont {A.}~\bibnamefont {Majumdar}},\ }\bibfield  {title}
  {\enquote {\bibinfo {title} {A review of heat transfer physics},}\ }\href
  {\doibase 10.1080/15567260801917520} {\bibfield  {journal} {\bibinfo
  {journal} {Nanoscale and Microscale Thermophysical Engineering}\ }\textbf
  {\bibinfo {volume} {12}},\ \bibinfo {pages} {1--60} (\bibinfo {year}
  {2008})},\ \Eprint
  {http://arxiv.org/abs/http://www.tandfonline.com/doi/pdf/10.1080/15567260801917520}
  {http://www.tandfonline.com/doi/pdf/10.1080/15567260801917520} \BibitemShut
  {NoStop}%
\bibitem [{\citenamefont {Donchev}(2006)}]{DonchevJCP2006}%
  \BibitemOpen
  \bibfield  {author} {\bibinfo {author} {\bibfnamefont {A.~G.}\ \bibnamefont
  {Donchev}},\ }\bibfield  {title} {\enquote {\bibinfo {title} {Many-body
  effects of dispersion interaction},}\ }\href {\doibase
  http://dx.doi.org/10.1063/1.2337283} {\bibfield  {journal} {\bibinfo
  {journal} {The Journal of Chemical Physics}\ }\textbf {\bibinfo {volume}
  {125}},\ \bibinfo {pages} {074713} (\bibinfo {year} {2006})}\BibitemShut
  {NoStop}%
\bibitem [{\citenamefont {Phan}\ \emph {et~al.}(2013)\citenamefont {Phan},
  \citenamefont {Woods},\ and\ \citenamefont {Phan}}]{PhanJAP2013}%
  \BibitemOpen
  \bibfield  {author} {\bibinfo {author} {\bibfnamefont {A.~D.}\ \bibnamefont
  {Phan}}, \bibinfo {author} {\bibfnamefont {L.~M.}\ \bibnamefont {Woods}}, \
  and\ \bibinfo {author} {\bibfnamefont {T.-L.}\ \bibnamefont {Phan}},\
  }\bibfield  {title} {\enquote {\bibinfo {title} {Van der waals interactions
  between graphitic nanowiggles},}\ }\href {\doibase
  http://dx.doi.org/10.1063/1.4816446} {\bibfield  {journal} {\bibinfo
  {journal} {Journal of Applied Physics}\ }\textbf {\bibinfo {volume} {114}}
  (\bibinfo {year} {2013}),\ http://dx.doi.org/10.1063/1.4816446}\BibitemShut
  {NoStop}%
\bibitem [{\citenamefont {Shtogun}\ and\ \citenamefont
  {Woods}(2010)}]{ShtogunJPCL2010}%
  \BibitemOpen
  \bibfield  {author} {\bibinfo {author} {\bibfnamefont {Y.~V.}\ \bibnamefont
  {Shtogun}}\ and\ \bibinfo {author} {\bibfnamefont {L.~M.}\ \bibnamefont
  {Woods}},\ }\bibfield  {title} {\enquote {\bibinfo {title} {Many-body van der
  waals interactions between graphitic nanostructures},}\ }\href {\doibase
  10.1021/jz100309m} {\bibfield  {journal} {\bibinfo  {journal} {The Journal of
  Physical Chemistry Letters}\ }\textbf {\bibinfo {volume} {1}},\ \bibinfo
  {pages} {1356--1362} (\bibinfo {year} {2010})}\BibitemShut {NoStop}%
\bibitem [{\citenamefont {Kim}\ \emph {et~al.}(2007)\citenamefont {Kim},
  \citenamefont {Sofo}, \citenamefont {Velegol}, \citenamefont {Cole},\ and\
  \citenamefont {Lucas}}]{KimLANGMUIR2007}%
  \BibitemOpen
  \bibfield  {author} {\bibinfo {author} {\bibfnamefont {H.-Y.}\ \bibnamefont
  {Kim}}, \bibinfo {author} {\bibfnamefont {J.~O.}\ \bibnamefont {Sofo}},
  \bibinfo {author} {\bibfnamefont {D.}~\bibnamefont {Velegol}}, \bibinfo
  {author} {\bibfnamefont {M.~W.}\ \bibnamefont {Cole}}, \ and\ \bibinfo
  {author} {\bibfnamefont {A.~A.}\ \bibnamefont {Lucas}},\ }\bibfield  {title}
  {\enquote {\bibinfo {title} {Van der waals dispersion forces between
  dielectric nanoclusters},}\ }\href {\doibase 10.1021/la061802w} {\bibfield
  {journal} {\bibinfo  {journal} {Langmuir}\ }\textbf {\bibinfo {volume}
  {23}},\ \bibinfo {pages} {1735--1740} (\bibinfo {year} {2007})}\BibitemShut
  {NoStop}%
\bibitem [{\citenamefont {Cole}\ \emph {et~al.}(2009)\citenamefont {Cole},
  \citenamefont {Velegol}, \citenamefont {Kim},\ and\ \citenamefont
  {Lucas}}]{ColeMS2009}%
  \BibitemOpen
  \bibfield  {author} {\bibinfo {author} {\bibfnamefont {M.~W.}\ \bibnamefont
  {Cole}}, \bibinfo {author} {\bibfnamefont {D.}~\bibnamefont {Velegol}},
  \bibinfo {author} {\bibfnamefont {H.-Y.}\ \bibnamefont {Kim}}, \ and\
  \bibinfo {author} {\bibfnamefont {A.~A.}\ \bibnamefont {Lucas}},\ }\bibfield
  {title} {\enquote {\bibinfo {title} {Nanoscale van der waals interactions},}\
  }\href {\doibase 10.1080/08927020902929794} {\bibfield  {journal} {\bibinfo
  {journal} {Molecular Simulation}\ }\textbf {\bibinfo {volume} {35}},\
  \bibinfo {pages} {849--866} (\bibinfo {year} {2009})}\BibitemShut {NoStop}%
\bibitem [{\citenamefont {Hermann}\ \emph {et~al.}(2017)\citenamefont
  {Hermann}, \citenamefont {DiStasio},\ and\ \citenamefont
  {Tkatchenko}}]{HermannCR2017}%
  \BibitemOpen
  \bibfield  {author} {\bibinfo {author} {\bibfnamefont {J.}~\bibnamefont
  {Hermann}}, \bibinfo {author} {\bibfnamefont {R.~A.}\ \bibnamefont
  {DiStasio}}, \ and\ \bibinfo {author} {\bibfnamefont {A.}~\bibnamefont
  {Tkatchenko}},\ }\bibfield  {title} {\enquote {\bibinfo {title}
  {First-principles models for van der waals interactions in molecules and
  materials: Concepts, theory, and applications},}\ }\href {\doibase
  10.1021/acs.chemrev.6b00446} {\bibfield  {journal} {\bibinfo  {journal}
  {Chemical Reviews}\ }\textbf {\bibinfo {volume} {117}},\ \bibinfo {pages}
  {4714--4758} (\bibinfo {year} {2017})},\ \bibinfo {note} {pMID: 28272886},\
  \Eprint {http://arxiv.org/abs/https://doi.org/10.1021/acs.chemrev.6b00446}
  {https://doi.org/10.1021/acs.chemrev.6b00446} \BibitemShut {NoStop}%
\bibitem [{\citenamefont {Wunsch}\ \emph {et~al.}(2006)\citenamefont {Wunsch},
  \citenamefont {Stauber}, \citenamefont {Sols},\ and\ \citenamefont
  {Guinea}}]{WunschNJP2006}%
  \BibitemOpen
  \bibfield  {author} {\bibinfo {author} {\bibfnamefont {B.}~\bibnamefont
  {Wunsch}}, \bibinfo {author} {\bibfnamefont {T.}~\bibnamefont {Stauber}},
  \bibinfo {author} {\bibfnamefont {F.}~\bibnamefont {Sols}}, \ and\ \bibinfo
  {author} {\bibfnamefont {F.}~\bibnamefont {Guinea}},\ }\bibfield  {title}
  {\enquote {\bibinfo {title} {Dynamical polarization of graphene at finite
  doping},}\ }\href {http://stacks.iop.org/1367-2630/8/i=12/a=318} {\bibfield
  {journal} {\bibinfo  {journal} {New Journal of Physics}\ }\textbf {\bibinfo
  {volume} {8}},\ \bibinfo {pages} {318} (\bibinfo {year} {2006})}\BibitemShut
  {NoStop}%
\bibitem [{\citenamefont {Wenger}\ \emph {et~al.}(2016)\citenamefont {Wenger},
  \citenamefont {Viola}, \citenamefont {Fogelstr\"om}, \citenamefont {Tassin},\
  and\ \citenamefont {Kinaret}}]{WengerPRB2016}%
  \BibitemOpen
  \bibfield  {author} {\bibinfo {author} {\bibfnamefont {T.}~\bibnamefont
  {Wenger}}, \bibinfo {author} {\bibfnamefont {G.}~\bibnamefont {Viola}},
  \bibinfo {author} {\bibfnamefont {M.}~\bibnamefont {Fogelstr\"om}}, \bibinfo
  {author} {\bibfnamefont {P.}~\bibnamefont {Tassin}}, \ and\ \bibinfo {author}
  {\bibfnamefont {J.}~\bibnamefont {Kinaret}},\ }\bibfield  {title} {\enquote
  {\bibinfo {title} {Optical signatures of nonlocal plasmons in graphene},}\
  }\href {\doibase 10.1103/PhysRevB.94.205419} {\bibfield  {journal} {\bibinfo
  {journal} {Phys. Rev. B}\ }\textbf {\bibinfo {volume} {94}},\ \bibinfo
  {pages} {205419} (\bibinfo {year} {2016})}\BibitemShut {NoStop}%
\bibitem [{\citenamefont {Dobson}(2011)}]{DobsonSS2011}%
  \BibitemOpen
  \bibfield  {author} {\bibinfo {author} {\bibfnamefont {J.~F.}\ \bibnamefont
  {Dobson}},\ }\bibfield  {title} {\enquote {\bibinfo {title} {Dispersion and
  induction interactions of graphene with nanostructures},}\ }\href {\doibase
  https://doi.org/10.1016/j.susc.2010.12.031} {\bibfield  {journal} {\bibinfo
  {journal} {Surface Science}\ }\textbf {\bibinfo {volume} {605}},\ \bibinfo
  {pages} {1621--1632} (\bibinfo {year} {2011})},\ \bibinfo {note} {graphene
  Surfaces and Interfaces}\BibitemShut {NoStop}%
\bibitem [{\citenamefont {Hwang}\ and\ \citenamefont
  {Das~Sarma}(2007)}]{HwangPRB2007}%
  \BibitemOpen
  \bibfield  {author} {\bibinfo {author} {\bibfnamefont {E.~H.}\ \bibnamefont
  {Hwang}}\ and\ \bibinfo {author} {\bibfnamefont {S.}~\bibnamefont
  {Das~Sarma}},\ }\bibfield  {title} {\enquote {\bibinfo {title} {Dielectric
  function, screening, and plasmons in two-dimensional graphene},}\ }\href
  {\doibase 10.1103/PhysRevB.75.205418} {\bibfield  {journal} {\bibinfo
  {journal} {Phys. Rev. B}\ }\textbf {\bibinfo {volume} {75}},\ \bibinfo
  {pages} {205418} (\bibinfo {year} {2007})}\BibitemShut {NoStop}%
\bibitem [{\citenamefont {Lundeberg}\ \emph {et~al.}(2017)\citenamefont
  {Lundeberg}, \citenamefont {Gao}, \citenamefont {Asgari}, \citenamefont
  {Tan}, \citenamefont {Van~Duppen}, \citenamefont {Autore}, \citenamefont
  {Alonso-Gonz{\'a}lez}, \citenamefont {Woessner}, \citenamefont {Watanabe},
  \citenamefont {Taniguchi}, \citenamefont {Hillenbrand}, \citenamefont {Hone},
  \citenamefont {Polini},\ and\ \citenamefont
  {Koppens}}]{LundebergSCIENCE2017}%
  \BibitemOpen
  \bibfield  {author} {\bibinfo {author} {\bibfnamefont {M.~B.}\ \bibnamefont
  {Lundeberg}}, \bibinfo {author} {\bibfnamefont {Y.}~\bibnamefont {Gao}},
  \bibinfo {author} {\bibfnamefont {R.}~\bibnamefont {Asgari}}, \bibinfo
  {author} {\bibfnamefont {C.}~\bibnamefont {Tan}}, \bibinfo {author}
  {\bibfnamefont {B.}~\bibnamefont {Van~Duppen}}, \bibinfo {author}
  {\bibfnamefont {M.}~\bibnamefont {Autore}}, \bibinfo {author} {\bibfnamefont
  {P.}~\bibnamefont {Alonso-Gonz{\'a}lez}}, \bibinfo {author} {\bibfnamefont
  {A.}~\bibnamefont {Woessner}}, \bibinfo {author} {\bibfnamefont
  {K.}~\bibnamefont {Watanabe}}, \bibinfo {author} {\bibfnamefont
  {T.}~\bibnamefont {Taniguchi}}, \bibinfo {author} {\bibfnamefont
  {R.}~\bibnamefont {Hillenbrand}}, \bibinfo {author} {\bibfnamefont
  {J.}~\bibnamefont {Hone}}, \bibinfo {author} {\bibfnamefont {M.}~\bibnamefont
  {Polini}}, \ and\ \bibinfo {author} {\bibfnamefont {F.~H.~L.}\ \bibnamefont
  {Koppens}},\ }\bibfield  {title} {\enquote {\bibinfo {title} {Tuning quantum
  nonlocal effects in graphene plasmonics},}\ }\href {\doibase
  10.1126/science.aan2735} {\bibfield  {journal} {\bibinfo  {journal}
  {Science}\ }\textbf {\bibinfo {volume} {357}},\ \bibinfo {pages} {187--191}
  (\bibinfo {year} {2017})},\ \Eprint
  {http://arxiv.org/abs/http://science.sciencemag.org/content/357/6347/187.full.pdf}
  {http://science.sciencemag.org/content/357/6347/187.full.pdf} \BibitemShut
  {NoStop}%
\bibitem [{\citenamefont {Sernelius}(2012)}]{SerneliusPRB2012}%
  \BibitemOpen
  \bibfield  {author} {\bibinfo {author} {\bibfnamefont {B.~E.}\ \bibnamefont
  {Sernelius}},\ }\bibfield  {title} {\enquote {\bibinfo {title} {Retarded
  interactions in graphene systems},}\ }\href {\doibase
  10.1103/PhysRevB.85.195427} {\bibfield  {journal} {\bibinfo  {journal} {Phys.
  Rev. B}\ }\textbf {\bibinfo {volume} {85}},\ \bibinfo {pages} {195427}
  (\bibinfo {year} {2012})}\BibitemShut {NoStop}%
\bibitem [{\citenamefont {Tsoi}\ \emph {et~al.}(2014)\citenamefont {Tsoi},
  \citenamefont {Dev}, \citenamefont {Friedman}, \citenamefont {Stine},
  \citenamefont {Robinson}, \citenamefont {Reinecke},\ and\ \citenamefont
  {Sheehan}}]{TsoiACSNANO2014}%
  \BibitemOpen
  \bibfield  {author} {\bibinfo {author} {\bibfnamefont {S.}~\bibnamefont
  {Tsoi}}, \bibinfo {author} {\bibfnamefont {P.}~\bibnamefont {Dev}}, \bibinfo
  {author} {\bibfnamefont {A.~L.}\ \bibnamefont {Friedman}}, \bibinfo {author}
  {\bibfnamefont {R.}~\bibnamefont {Stine}}, \bibinfo {author} {\bibfnamefont
  {J.~T.}\ \bibnamefont {Robinson}}, \bibinfo {author} {\bibfnamefont {T.~L.}\
  \bibnamefont {Reinecke}}, \ and\ \bibinfo {author} {\bibfnamefont {P.~E.}\
  \bibnamefont {Sheehan}},\ }\bibfield  {title} {\enquote {\bibinfo {title}
  {van der waals screening by single-layer graphene and molybdenum
  disulfide},}\ }\href {\doibase 10.1021/nn5050905} {\bibfield  {journal}
  {\bibinfo  {journal} {ACS Nano}\ }\textbf {\bibinfo {volume} {8}},\ \bibinfo
  {pages} {12410--12417} (\bibinfo {year} {2014})},\ \bibinfo {note} {pMID:
  25412420},\ \Eprint {http://arxiv.org/abs/https://doi.org/10.1021/nn5050905}
  {https://doi.org/10.1021/nn5050905} \BibitemShut {NoStop}%
\bibitem [{\citenamefont {Tang}\ \emph {et~al.}(2017)\citenamefont {Tang},
  \citenamefont {Wang}, \citenamefont {Ng}, \citenamefont {Nikolic},
  \citenamefont {Chan}, \citenamefont {Rodriguez},\ and\ \citenamefont
  {Chan}}]{TangNATURE2017}%
  \BibitemOpen
  \bibfield  {author} {\bibinfo {author} {\bibfnamefont {L.}~\bibnamefont
  {Tang}}, \bibinfo {author} {\bibfnamefont {M.}~\bibnamefont {Wang}}, \bibinfo
  {author} {\bibfnamefont {C.}~\bibnamefont {Ng}}, \bibinfo {author}
  {\bibfnamefont {M.}~\bibnamefont {Nikolic}}, \bibinfo {author} {\bibfnamefont
  {C.~T.}\ \bibnamefont {Chan}}, \bibinfo {author} {\bibfnamefont {A.~W.}\
  \bibnamefont {Rodriguez}}, \ and\ \bibinfo {author} {\bibfnamefont {H.~B.}\
  \bibnamefont {Chan}},\ }\bibfield  {title} {\enquote {\bibinfo {title}
  {Measurement of non-monotonic casimir forces between silicon
  nanostructures},}\ }\href@noop {} {\bibfield  {journal} {\bibinfo  {journal}
  {Nature Photonics}\ }\textbf {\bibinfo {volume} {11}},\ \bibinfo {pages} {97}
  (\bibinfo {year} {2017})}\BibitemShut {NoStop}%
\bibitem [{\citenamefont {Garrett}\ \emph {et~al.}(2018)\citenamefont
  {Garrett}, \citenamefont {Somers},\ and\ \citenamefont
  {Munday}}]{GarrettPRL2018}%
  \BibitemOpen
  \bibfield  {author} {\bibinfo {author} {\bibfnamefont {J.~L.}\ \bibnamefont
  {Garrett}}, \bibinfo {author} {\bibfnamefont {D.~A.~T.}\ \bibnamefont
  {Somers}}, \ and\ \bibinfo {author} {\bibfnamefont {J.~N.}\ \bibnamefont
  {Munday}},\ }\bibfield  {title} {\enquote {\bibinfo {title} {Measurement of
  the casimir force between two spheres},}\ }\href {\doibase
  10.1103/PhysRevLett.120.040401} {\bibfield  {journal} {\bibinfo  {journal}
  {Phys. Rev. Lett.}\ }\textbf {\bibinfo {volume} {120}},\ \bibinfo {pages}
  {040401} (\bibinfo {year} {2018})}\BibitemShut {NoStop}%
\bibitem [{\citenamefont {Maghrebi}\ \emph {et~al.}(2011)\citenamefont
  {Maghrebi}, \citenamefont {Rahi}, \citenamefont {Emig}, \citenamefont
  {Graham}, \citenamefont {Jaffe},\ and\ \citenamefont
  {Kardar}}]{MaghrebiPNAS2011}%
  \BibitemOpen
  \bibfield  {author} {\bibinfo {author} {\bibfnamefont {M.~F.}\ \bibnamefont
  {Maghrebi}}, \bibinfo {author} {\bibfnamefont {S.~J.}\ \bibnamefont {Rahi}},
  \bibinfo {author} {\bibfnamefont {T.}~\bibnamefont {Emig}}, \bibinfo {author}
  {\bibfnamefont {N.}~\bibnamefont {Graham}}, \bibinfo {author} {\bibfnamefont
  {R.~L.}\ \bibnamefont {Jaffe}}, \ and\ \bibinfo {author} {\bibfnamefont
  {M.}~\bibnamefont {Kardar}},\ }\bibfield  {title} {\enquote {\bibinfo {title}
  {Analytical results on casimir forces for conductors with edges and tips},}\
  }\href {\doibase 10.1073/pnas.1018079108} {\bibfield  {journal} {\bibinfo
  {journal} {Proceedings of the National Academy of Sciences}\ }\textbf
  {\bibinfo {volume} {108}},\ \bibinfo {pages} {6867--6871} (\bibinfo {year}
  {2011})},\ \Eprint
  {http://arxiv.org/abs/http://www.pnas.org/content/108/17/6867.full.pdf}
  {http://www.pnas.org/content/108/17/6867.full.pdf} \BibitemShut {NoStop}%
\bibitem [{\citenamefont {Sernelius}(2006)}]{SerneliusJPA2006}%
  \BibitemOpen
  \bibfield  {author} {\bibinfo {author} {\bibfnamefont {B.~E.}\ \bibnamefont
  {Sernelius}},\ }\bibfield  {title} {\enquote {\bibinfo {title}
  {Finite-temperature casimir force between metal plates: full inclusion of
  spatial dispersion resolves a long-standing controversy},}\ }\href
  {http://stacks.iop.org/0305-4470/39/i=21/a=S75} {\bibfield  {journal}
  {\bibinfo  {journal} {Journal of Physics A: Mathematical and General}\
  }\textbf {\bibinfo {volume} {39}},\ \bibinfo {pages} {6741} (\bibinfo {year}
  {2006})}\BibitemShut {NoStop}%
\bibitem [{\citenamefont {Chalifoux}\ and\ \citenamefont
  {Tykwinski}(2010)}]{ChalifouxNATURE2010}%
  \BibitemOpen
  \bibfield  {author} {\bibinfo {author} {\bibfnamefont {W.~A.}\ \bibnamefont
  {Chalifoux}}\ and\ \bibinfo {author} {\bibfnamefont {R.~R.}\ \bibnamefont
  {Tykwinski}},\ }\bibfield  {title} {\enquote {\bibinfo {title} {Synthesis of
  polyynes to model the sp-carbon allotrope carbyne},}\ }\href@noop {}
  {\bibfield  {journal} {\bibinfo  {journal} {Nature chemistry}\ }\textbf
  {\bibinfo {volume} {2}},\ \bibinfo {pages} {967--971} (\bibinfo {year}
  {2010})}\BibitemShut {NoStop}%
\bibitem [{\citenamefont {Shi}\ \emph {et~al.}(2016)\citenamefont {Shi},
  \citenamefont {Rohringer}, \citenamefont {Suenaga}, \citenamefont {Niimi},
  \citenamefont {Kotakoski}, \citenamefont {Meyer}, \citenamefont {Peterlik},
  \citenamefont {Wanko}, \citenamefont {Cahangirov}, \citenamefont {Rubio}
  \emph {et~al.}}]{ShiNATURE2016}%
  \BibitemOpen
  \bibfield  {author} {\bibinfo {author} {\bibfnamefont {L.}~\bibnamefont
  {Shi}}, \bibinfo {author} {\bibfnamefont {P.}~\bibnamefont {Rohringer}},
  \bibinfo {author} {\bibfnamefont {K.}~\bibnamefont {Suenaga}}, \bibinfo
  {author} {\bibfnamefont {Y.}~\bibnamefont {Niimi}}, \bibinfo {author}
  {\bibfnamefont {J.}~\bibnamefont {Kotakoski}}, \bibinfo {author}
  {\bibfnamefont {J.~C.}\ \bibnamefont {Meyer}}, \bibinfo {author}
  {\bibfnamefont {H.}~\bibnamefont {Peterlik}}, \bibinfo {author}
  {\bibfnamefont {M.}~\bibnamefont {Wanko}}, \bibinfo {author} {\bibfnamefont
  {S.}~\bibnamefont {Cahangirov}}, \bibinfo {author} {\bibfnamefont
  {A.}~\bibnamefont {Rubio}},  \emph {et~al.},\ }\bibfield  {title} {\enquote
  {\bibinfo {title} {Confined linear carbon chains as a route to bulk
  carbyne},}\ }\href@noop {} {\bibfield  {journal} {\bibinfo  {journal} {Nature
  materials}\ }\textbf {\bibinfo {volume} {15}},\ \bibinfo {pages} {634}
  (\bibinfo {year} {2016})}\BibitemShut {NoStop}%
\bibitem [{\citenamefont {Yi}\ and\ \citenamefont {Shen}(2015)}]{YiJMCA2015}%
  \BibitemOpen
  \bibfield  {author} {\bibinfo {author} {\bibfnamefont {M.}~\bibnamefont
  {Yi}}\ and\ \bibinfo {author} {\bibfnamefont {Z.}~\bibnamefont {Shen}},\
  }\bibfield  {title} {\enquote {\bibinfo {title} {A review on mechanical
  exfoliation for the scalable production of graphene},}\ }\href {\doibase
  10.1039/C5TA00252D} {\bibfield  {journal} {\bibinfo  {journal} {J. Mater.
  Chem. A}\ }\textbf {\bibinfo {volume} {3}},\ \bibinfo {pages} {11700--11715}
  (\bibinfo {year} {2015})}\BibitemShut {NoStop}%
\bibitem [{\citenamefont {Bunch}\ \emph {et~al.}(2007)\citenamefont {Bunch},
  \citenamefont {van~der Zande}, \citenamefont {Verbridge}, \citenamefont
  {Frank}, \citenamefont {Tanenbaum}, \citenamefont {Parpia}, \citenamefont
  {Craighead},\ and\ \citenamefont {McEuen}}]{BunchSCIENCE2007}%
  \BibitemOpen
  \bibfield  {author} {\bibinfo {author} {\bibfnamefont {J.~S.}\ \bibnamefont
  {Bunch}}, \bibinfo {author} {\bibfnamefont {A.~M.}\ \bibnamefont {van~der
  Zande}}, \bibinfo {author} {\bibfnamefont {S.~S.}\ \bibnamefont {Verbridge}},
  \bibinfo {author} {\bibfnamefont {I.~W.}\ \bibnamefont {Frank}}, \bibinfo
  {author} {\bibfnamefont {D.~M.}\ \bibnamefont {Tanenbaum}}, \bibinfo {author}
  {\bibfnamefont {J.~M.}\ \bibnamefont {Parpia}}, \bibinfo {author}
  {\bibfnamefont {H.~G.}\ \bibnamefont {Craighead}}, \ and\ \bibinfo {author}
  {\bibfnamefont {P.~L.}\ \bibnamefont {McEuen}},\ }\bibfield  {title}
  {\enquote {\bibinfo {title} {Electromechanical resonators from graphene
  sheets},}\ }\href {\doibase 10.1126/science.1136836} {\bibfield  {journal}
  {\bibinfo  {journal} {Science}\ }\textbf {\bibinfo {volume} {315}},\ \bibinfo
  {pages} {490--493} (\bibinfo {year} {2007})},\ \Eprint
  {http://arxiv.org/abs/http://science.sciencemag.org/content/315/5811/490.full.pdf}
  {http://science.sciencemag.org/content/315/5811/490.full.pdf} \BibitemShut
  {NoStop}%
\bibitem [{\citenamefont {Falkovsky}(2007)}]{FalkovskyJETP2007}%
  \BibitemOpen
  \bibfield  {author} {\bibinfo {author} {\bibfnamefont {L.~A.}\ \bibnamefont
  {Falkovsky}},\ }\bibfield  {title} {\enquote {\bibinfo {title} {Phonon
  dispersion in graphene},}\ }\href {\doibase 10.1134/S1063776107080122}
  {\bibfield  {journal} {\bibinfo  {journal} {Journal of Experimental and
  Theoretical Physics}\ }\textbf {\bibinfo {volume} {105}},\ \bibinfo {pages}
  {397--403} (\bibinfo {year} {2007})}\BibitemShut {NoStop}%
\bibitem [{\citenamefont {Endlich}\ \emph {et~al.}(2013)\citenamefont
  {Endlich}, \citenamefont {Molina-S\'anchez}, \citenamefont {Wirtz},\ and\
  \citenamefont {Kr\"oger}}]{EndlichPRB2013}%
  \BibitemOpen
  \bibfield  {author} {\bibinfo {author} {\bibfnamefont {M.}~\bibnamefont
  {Endlich}}, \bibinfo {author} {\bibfnamefont {A.}~\bibnamefont
  {Molina-S\'anchez}}, \bibinfo {author} {\bibfnamefont {L.}~\bibnamefont
  {Wirtz}}, \ and\ \bibinfo {author} {\bibfnamefont {J.}~\bibnamefont
  {Kr\"oger}},\ }\bibfield  {title} {\enquote {\bibinfo {title} {Screening of
  electron-phonon coupling in graphene on ir(111)},}\ }\href {\doibase
  10.1103/PhysRevB.88.205403} {\bibfield  {journal} {\bibinfo  {journal} {Phys.
  Rev. B}\ }\textbf {\bibinfo {volume} {88}},\ \bibinfo {pages} {205403}
  (\bibinfo {year} {2013})}\BibitemShut {NoStop}%
\bibitem [{\citenamefont {Akinwande}\ \emph {et~al.}(2017)\citenamefont
  {Akinwande}, \citenamefont {Brennan}, \citenamefont {Bunch}, \citenamefont
  {Egberts}, \citenamefont {Felts}, \citenamefont {Gao}, \citenamefont {Huang},
  \citenamefont {Kim}, \citenamefont {Li}, \citenamefont {Li}, \citenamefont
  {Liechti}, \citenamefont {Lu}, \citenamefont {Park}, \citenamefont {Reed},
  \citenamefont {Wang}, \citenamefont {Yakobson}, \citenamefont {Zhang},
  \citenamefont {Zhang}, \citenamefont {Zhou},\ and\ \citenamefont
  {Zhu}}]{AkinwandeEML2017}%
  \BibitemOpen
  \bibfield  {author} {\bibinfo {author} {\bibfnamefont {D.}~\bibnamefont
  {Akinwande}}, \bibinfo {author} {\bibfnamefont {C.~J.}\ \bibnamefont
  {Brennan}}, \bibinfo {author} {\bibfnamefont {J.~S.}\ \bibnamefont {Bunch}},
  \bibinfo {author} {\bibfnamefont {P.}~\bibnamefont {Egberts}}, \bibinfo
  {author} {\bibfnamefont {J.~R.}\ \bibnamefont {Felts}}, \bibinfo {author}
  {\bibfnamefont {H.}~\bibnamefont {Gao}}, \bibinfo {author} {\bibfnamefont
  {R.}~\bibnamefont {Huang}}, \bibinfo {author} {\bibfnamefont {J.-S.}\
  \bibnamefont {Kim}}, \bibinfo {author} {\bibfnamefont {T.}~\bibnamefont
  {Li}}, \bibinfo {author} {\bibfnamefont {Y.}~\bibnamefont {Li}}, \bibinfo
  {author} {\bibfnamefont {K.~M.}\ \bibnamefont {Liechti}}, \bibinfo {author}
  {\bibfnamefont {N.}~\bibnamefont {Lu}}, \bibinfo {author} {\bibfnamefont
  {H.~S.}\ \bibnamefont {Park}}, \bibinfo {author} {\bibfnamefont {E.~J.}\
  \bibnamefont {Reed}}, \bibinfo {author} {\bibfnamefont {P.}~\bibnamefont
  {Wang}}, \bibinfo {author} {\bibfnamefont {B.~I.}\ \bibnamefont {Yakobson}},
  \bibinfo {author} {\bibfnamefont {T.}~\bibnamefont {Zhang}}, \bibinfo
  {author} {\bibfnamefont {Y.-W.}\ \bibnamefont {Zhang}}, \bibinfo {author}
  {\bibfnamefont {Y.}~\bibnamefont {Zhou}}, \ and\ \bibinfo {author}
  {\bibfnamefont {Y.}~\bibnamefont {Zhu}},\ }\bibfield  {title} {\enquote
  {\bibinfo {title} {A review on mechanics and mechanical properties of 2d
  materials--graphene and beyond},}\ }\href {\doibase
  https://doi.org/10.1016/j.eml.2017.01.008} {\bibfield  {journal} {\bibinfo
  {journal} {Extreme Mechanics Letters}\ }\textbf {\bibinfo {volume} {13}},\
  \bibinfo {pages} {42 -- 77} (\bibinfo {year} {2017})}\BibitemShut {NoStop}%
\bibitem [{\citenamefont {Castro~Neto}\ \emph {et~al.}(2009)\citenamefont
  {Castro~Neto}, \citenamefont {Guinea}, \citenamefont {Peres}, \citenamefont
  {Novoselov},\ and\ \citenamefont {Geim}}]{CastroNetoRMP2009}%
  \BibitemOpen
  \bibfield  {author} {\bibinfo {author} {\bibfnamefont {A.~H.}\ \bibnamefont
  {Castro~Neto}}, \bibinfo {author} {\bibfnamefont {F.}~\bibnamefont {Guinea}},
  \bibinfo {author} {\bibfnamefont {N.~M.~R.}\ \bibnamefont {Peres}}, \bibinfo
  {author} {\bibfnamefont {K.~S.}\ \bibnamefont {Novoselov}}, \ and\ \bibinfo
  {author} {\bibfnamefont {A.~K.}\ \bibnamefont {Geim}},\ }\bibfield  {title}
  {\enquote {\bibinfo {title} {The electronic properties of graphene},}\ }\href
  {\doibase 10.1103/RevModPhys.81.109} {\bibfield  {journal} {\bibinfo
  {journal} {Rev. Mod. Phys.}\ }\textbf {\bibinfo {volume} {81}},\ \bibinfo
  {pages} {109--162} (\bibinfo {year} {2009})}\BibitemShut {NoStop}%
\bibitem [{\citenamefont {Balandin}(2011)}]{BalandinNATURE2011}%
  \BibitemOpen
  \bibfield  {author} {\bibinfo {author} {\bibfnamefont {A.~A.}\ \bibnamefont
  {Balandin}},\ }\bibfield  {title} {\enquote {\bibinfo {title} {Thermal
  properties of graphene and nanostructured carbon materials},}\ }\href@noop {}
  {\bibfield  {journal} {\bibinfo  {journal} {Nature materials}\ }\textbf
  {\bibinfo {volume} {10}},\ \bibinfo {pages} {569} (\bibinfo {year}
  {2011})}\BibitemShut {NoStop}%
\bibitem [{\citenamefont {Li}\ and\ \citenamefont {Han}(2018)}]{LiPRB2018}%
  \BibitemOpen
  \bibfield  {author} {\bibinfo {author} {\bibfnamefont {J.}~\bibnamefont
  {Li}}\ and\ \bibinfo {author} {\bibfnamefont {J.~E.}\ \bibnamefont {Han}},\
  }\bibfield  {title} {\enquote {\bibinfo {title} {Nonequilibrium excitations
  and transport of dirac electrons in electric-field-driven graphene},}\ }\href
  {\doibase 10.1103/PhysRevB.97.205412} {\bibfield  {journal} {\bibinfo
  {journal} {Phys. Rev. B}\ }\textbf {\bibinfo {volume} {97}},\ \bibinfo
  {pages} {205412} (\bibinfo {year} {2018})}\BibitemShut {NoStop}%
\bibitem [{\citenamefont {Ilic}\ \emph {et~al.}(2012)\citenamefont {Ilic},
  \citenamefont {Jablan}, \citenamefont {Joannopoulos}, \citenamefont
  {Celanovic}, \citenamefont {Buljan},\ and\ \citenamefont {Solja\ifmmode
  \check{c}\else \v{c}\fi{}i\ifmmode~\acute{c}\else \'{c}\fi{}}}]{IlicPRB2012}%
  \BibitemOpen
  \bibfield  {author} {\bibinfo {author} {\bibfnamefont {O.}~\bibnamefont
  {Ilic}}, \bibinfo {author} {\bibfnamefont {M.}~\bibnamefont {Jablan}},
  \bibinfo {author} {\bibfnamefont {J.~D.}\ \bibnamefont {Joannopoulos}},
  \bibinfo {author} {\bibfnamefont {I.}~\bibnamefont {Celanovic}}, \bibinfo
  {author} {\bibfnamefont {H.}~\bibnamefont {Buljan}}, \ and\ \bibinfo {author}
  {\bibfnamefont {M.}~\bibnamefont {Solja\ifmmode \check{c}\else
  \v{c}\fi{}i\ifmmode~\acute{c}\else \'{c}\fi{}}},\ }\bibfield  {title}
  {\enquote {\bibinfo {title} {Near-field thermal radiation transfer controlled
  by plasmons in graphene},}\ }\href {\doibase 10.1103/PhysRevB.85.155422}
  {\bibfield  {journal} {\bibinfo  {journal} {Phys. Rev. B}\ }\textbf {\bibinfo
  {volume} {85}},\ \bibinfo {pages} {155422} (\bibinfo {year}
  {2012})}\BibitemShut {NoStop}%
\bibitem [{\citenamefont {Banishev}\ \emph {et~al.}(2013)\citenamefont
  {Banishev}, \citenamefont {Wen}, \citenamefont {Xu}, \citenamefont
  {Kawakami}, \citenamefont {Klimchitskaya}, \citenamefont {Mostepanenko},\
  and\ \citenamefont {Mohideen}}]{BanishevPRB2013}%
  \BibitemOpen
  \bibfield  {author} {\bibinfo {author} {\bibfnamefont {A.~A.}\ \bibnamefont
  {Banishev}}, \bibinfo {author} {\bibfnamefont {H.}~\bibnamefont {Wen}},
  \bibinfo {author} {\bibfnamefont {J.}~\bibnamefont {Xu}}, \bibinfo {author}
  {\bibfnamefont {R.~K.}\ \bibnamefont {Kawakami}}, \bibinfo {author}
  {\bibfnamefont {G.~L.}\ \bibnamefont {Klimchitskaya}}, \bibinfo {author}
  {\bibfnamefont {V.~M.}\ \bibnamefont {Mostepanenko}}, \ and\ \bibinfo
  {author} {\bibfnamefont {U.}~\bibnamefont {Mohideen}},\ }\bibfield  {title}
  {\enquote {\bibinfo {title} {Measuring the casimir force gradient from
  graphene on a sio${}_{2}$ substrate},}\ }\href {\doibase
  10.1103/PhysRevB.87.205433} {\bibfield  {journal} {\bibinfo  {journal} {Phys.
  Rev. B}\ }\textbf {\bibinfo {volume} {87}},\ \bibinfo {pages} {205433}
  (\bibinfo {year} {2013})}\BibitemShut {NoStop}%
\bibitem [{\citenamefont {Klimchitskaya}\ and\ \citenamefont
  {Mostepanenko}(2015)}]{KlimchitskayaPRB2015}%
  \BibitemOpen
  \bibfield  {author} {\bibinfo {author} {\bibfnamefont {G.~L.}\ \bibnamefont
  {Klimchitskaya}}\ and\ \bibinfo {author} {\bibfnamefont {V.~M.}\ \bibnamefont
  {Mostepanenko}},\ }\bibfield  {title} {\enquote {\bibinfo {title} {Comparison
  of hydrodynamic model of graphene with recent experiment on measuring the
  casimir interaction},}\ }\href {\doibase 10.1103/PhysRevB.91.045412}
  {\bibfield  {journal} {\bibinfo  {journal} {Phys. Rev. B}\ }\textbf {\bibinfo
  {volume} {91}},\ \bibinfo {pages} {045412} (\bibinfo {year}
  {2015})}\BibitemShut {NoStop}%
\bibitem [{\citenamefont {Jablan}\ \emph {et~al.}(2011)\citenamefont {Jablan},
  \citenamefont {Solja\ifmmode \check{c}\else
  \v{c}\fi{}i\ifmmode~\acute{c}\else \'{c}\fi{}},\ and\ \citenamefont
  {Buljan}}]{JablanPRB2011}%
  \BibitemOpen
  \bibfield  {author} {\bibinfo {author} {\bibfnamefont {M.}~\bibnamefont
  {Jablan}}, \bibinfo {author} {\bibfnamefont {M.}~\bibnamefont {Solja\ifmmode
  \check{c}\else \v{c}\fi{}i\ifmmode~\acute{c}\else \'{c}\fi{}}}, \ and\
  \bibinfo {author} {\bibfnamefont {H.}~\bibnamefont {Buljan}},\ }\bibfield
  {title} {\enquote {\bibinfo {title} {Unconventional plasmon-phonon coupling
  in graphene},}\ }\href {\doibase 10.1103/PhysRevB.83.161409} {\bibfield
  {journal} {\bibinfo  {journal} {Phys. Rev. B}\ }\textbf {\bibinfo {volume}
  {83}},\ \bibinfo {pages} {161409} (\bibinfo {year} {2011})}\BibitemShut
  {NoStop}%
\bibitem [{\citenamefont {Barlas}\ \emph {et~al.}(2007)\citenamefont {Barlas},
  \citenamefont {Pereg-Barnea}, \citenamefont {Polini}, \citenamefont
  {Asgari},\ and\ \citenamefont {MacDonald}}]{BarlasPRL2007}%
  \BibitemOpen
  \bibfield  {author} {\bibinfo {author} {\bibfnamefont {Y.}~\bibnamefont
  {Barlas}}, \bibinfo {author} {\bibfnamefont {T.}~\bibnamefont
  {Pereg-Barnea}}, \bibinfo {author} {\bibfnamefont {M.}~\bibnamefont
  {Polini}}, \bibinfo {author} {\bibfnamefont {R.}~\bibnamefont {Asgari}}, \
  and\ \bibinfo {author} {\bibfnamefont {A.~H.}\ \bibnamefont {MacDonald}},\
  }\bibfield  {title} {\enquote {\bibinfo {title} {Chirality and correlations
  in graphene},}\ }\href {\doibase 10.1103/PhysRevLett.98.236601} {\bibfield
  {journal} {\bibinfo  {journal} {Phys. Rev. Lett.}\ }\textbf {\bibinfo
  {volume} {98}},\ \bibinfo {pages} {236601} (\bibinfo {year}
  {2007})}\BibitemShut {NoStop}%
\bibitem [{\citenamefont {Bordag}\ \emph {et~al.}(2009)\citenamefont {Bordag},
  \citenamefont {Fialkovsky}, \citenamefont {Gitman},\ and\ \citenamefont
  {Vassilevich}}]{BordagPRB2009}%
  \BibitemOpen
  \bibfield  {author} {\bibinfo {author} {\bibfnamefont {M.}~\bibnamefont
  {Bordag}}, \bibinfo {author} {\bibfnamefont {I.~V.}\ \bibnamefont
  {Fialkovsky}}, \bibinfo {author} {\bibfnamefont {D.~M.}\ \bibnamefont
  {Gitman}}, \ and\ \bibinfo {author} {\bibfnamefont {D.~V.}\ \bibnamefont
  {Vassilevich}},\ }\bibfield  {title} {\enquote {\bibinfo {title} {Casimir
  interaction between a perfect conductor and graphene described by the dirac
  model},}\ }\href {\doibase 10.1103/PhysRevB.80.245406} {\bibfield  {journal}
  {\bibinfo  {journal} {Phys. Rev. B}\ }\textbf {\bibinfo {volume} {80}},\
  \bibinfo {pages} {245406} (\bibinfo {year} {2009})}\BibitemShut {NoStop}%
\bibitem [{\citenamefont {Bordag}\ \emph {et~al.}(2016)\citenamefont {Bordag},
  \citenamefont {Fialkovskiy},\ and\ \citenamefont
  {Vassilevich}}]{BordagPRB2016}%
  \BibitemOpen
  \bibfield  {author} {\bibinfo {author} {\bibfnamefont {M.}~\bibnamefont
  {Bordag}}, \bibinfo {author} {\bibfnamefont {I.}~\bibnamefont {Fialkovskiy}},
  \ and\ \bibinfo {author} {\bibfnamefont {D.}~\bibnamefont {Vassilevich}},\
  }\bibfield  {title} {\enquote {\bibinfo {title} {Enhanced casimir effect for
  doped graphene},}\ }\href {\doibase 10.1103/PhysRevB.93.075414} {\bibfield
  {journal} {\bibinfo  {journal} {Phys. Rev. B}\ }\textbf {\bibinfo {volume}
  {93}},\ \bibinfo {pages} {075414} (\bibinfo {year} {2016})}\BibitemShut
  {NoStop}%
\bibitem [{\citenamefont {Drosdoff}\ and\ \citenamefont
  {Woods}(2010)}]{DrosdoffPRB2010}%
  \BibitemOpen
  \bibfield  {author} {\bibinfo {author} {\bibfnamefont {D.}~\bibnamefont
  {Drosdoff}}\ and\ \bibinfo {author} {\bibfnamefont {L.~M.}\ \bibnamefont
  {Woods}},\ }\bibfield  {title} {\enquote {\bibinfo {title} {Casimir forces
  and graphene sheets},}\ }\href {\doibase 10.1103/PhysRevB.82.155459}
  {\bibfield  {journal} {\bibinfo  {journal} {Phys. Rev. B}\ }\textbf {\bibinfo
  {volume} {82}},\ \bibinfo {pages} {155459} (\bibinfo {year}
  {2010})}\BibitemShut {NoStop}%
\bibitem [{\citenamefont {Koppens}\ \emph {et~al.}(2011)\citenamefont
  {Koppens}, \citenamefont {Chang},\ and\ \citenamefont {Garc\'{i}a~de
  Abajo}}]{KoppensNANOLETT2011}%
  \BibitemOpen
  \bibfield  {author} {\bibinfo {author} {\bibfnamefont {F.~H.~L.}\
  \bibnamefont {Koppens}}, \bibinfo {author} {\bibfnamefont {D.~E.}\
  \bibnamefont {Chang}}, \ and\ \bibinfo {author} {\bibfnamefont {F.~J.}\
  \bibnamefont {Garc\'{i}a~de Abajo}},\ }\bibfield  {title} {\enquote {\bibinfo
  {title} {Graphene plasmonics: A platform for strong light--matter
  interactions},}\ }\href {\doibase 10.1021/nl201771h} {\bibfield  {journal}
  {\bibinfo  {journal} {Nano Letters}\ }\textbf {\bibinfo {volume} {11}},\
  \bibinfo {pages} {3370--3377} (\bibinfo {year} {2011})},\ \bibinfo {note}
  {pMID: 21766812},\ \Eprint
  {http://arxiv.org/abs/https://doi.org/10.1021/nl201771h}
  {https://doi.org/10.1021/nl201771h} \BibitemShut {NoStop}%
\end{thebibliography}%
\end{document}